\documentclass[a4paper,11pt]{article}
\usepackage{jcappub} 
\usepackage{mathtools}
\usepackage{booktabs}
\usepackage[english]{babel}
\usepackage{amsmath,amssymb,amsbsy,amstext, amsthm, simplewick, amsfonts}
\usepackage{graphicx}
\usepackage[small]{caption}
\usepackage{siunitx}
\usepackage{upgreek}
\usepackage{framed}
\usepackage{cleveref}
\usepackage{wrapfig}
\usepackage{multirow}
\usepackage{bbm}
\usepackage[svgnames,dvipsnames,x11names]{xcolor}
\usepackage[utf8x]{inputenc}
\usepackage{selinput}
\usepackage{bm}
\usepackage{float}
\usepackage{geometry}
\usepackage{yfonts}
\usepackage{caption}
\usepackage{subcaption}
\usepackage{subfloat}
\usepackage{sidecap}
\usepackage{longtable}
\usepackage{physics}
\usepackage{anyfontsize}
\usepackage{dsfont}
\usepackage{colortbl}
\usepackage{lineno}
\usepackage{comment}

\newcommand{\rom}[1]{\uppercase\expandafter{\romannumeral #1\relax}}

\newcommand{\calR}{{\mathcal R}}

\arxivnumber{YITP-24-174} 
\title{A complete analysis of inflation with piecewise quadratic potential}

\author{Xinpeng Wang$^{1,2}$, Xiao-Han Ma$^{1,3,4}$, Misao Sasaki$^{1,5,6}$}
\affiliation{
$^1$Kavli Institute for the Physics and Mathematics of the Universe (WPI), The University of Tokyo Institutes for Advanced Study, The University of Tokyo, Chiba 277-8583, Japan\\
$^2$ School of Physics Science and Engineering, Tongji University, Shanghai 200092, China\\
$^3$Deep Space Exploration Laboratory/School of Physical Sciences, University of Science and Technology of China, Hefei, Anhui 230026, China\\
$^4$CAS Key Laboratory for Researches in Galaxies and Cosmology/Department of Astronomy, School of Astronomy and Space Science, University of Science and Technology of China, Hefei, Anhui 230026, China\\
$^5$Center for Gravitational Physics and Quantum Information,
Yukawa Institute for Theoretical Physics, Kyoto University, Kyoto 606-8502, Japan\\
$^6$Leung Center for Cosmology and Particle Astrophysics,
National Taiwan University, Taipei 10617, Taiwan}

\emailAdd{xinpengwang@tongji.edu.cn, mxh171554@mail.ustc.edu.cn, misao.sasaki@ipmu.jp}
\abstract{
   We conduct a thorough study of the comoving curvature perturbation $\mathcal{R}$ in single-field inflation with two stages, represented by a piecewise quadratic potential, where both the first and second derivatives are allowed to be discontinuous at the transition point.
   We calculate the evolution of $\mathcal{R}$ by combining the perturbative and non-perturbative methods consistently, and obtain the power spectrum and the non-Gaussian features in the probability distribution function. 
   We find that both the spectrum and the statistics of $\mathcal{R}$ depend significantly on the second derivatives of the potential at both the first and second stages. 
   Furthermore, we find a new parameter constructed from the potential parameters, which we call $\alpha$, plays a decisive role in determining various features in the spectrum such as the amplitude, the slope, and the existence of a dip.
  In particular, we recover the typical $k^4$ growth of the spectrum in most cases, but the maximum growth rate of $k^5(\log k)^2$ can be obtained by fine-tuning the parameters.
Then, using the $\delta N$ formalism valid on superhorizon scales, we give fully nonlinear formulas for $\mathcal{R}$ in terms of the scalar field perturbation $\delta\phi$ and its time derivative. 
In passing, we point out the importance of the nonlinear evolution of $\delta\phi$ on superhorizon scales.
Finally, using the Press-Schechter formalism for simplicity, we discuss the effect of the non-Gaussian tails of the probability distribution function on the primordial black hole formation.
}

\begin{document}
\maketitle

\flushbottom

\section{Introduction}

Inflation, a quasi-exponential expansion stage before the radiation-dominated era, serves as a simple but effective phenomenological description of the very early universe. It was originally meant to solve several problems in the Big Bang cosmology, such as the initial singularity, horizon and flatness problems~\citep{Guth:1980zm, Starobinsky:1980te, Linde:1981mu}, though it is well-understood now that the most important prediction of inflation is that it explains the origin of all the structures in the universe. 
During inflation, the quantum vacuum fluctuations of a scalar field on microscopic scales are stretched beyond the Hubble radius and frozen on supper Hubble scales. After the end of inflation, these fluctuations eventually turn into curvature perturbation and provide seeds for the inhomogeneous structures in the universe today. \\

Based on the concordance $\Lambda$CDM model, the Planck satellite \cite{Planck:2018jri} has put tight constraints on primordial perturbations on large scales, indicating they are nearly scale-invariant, adiabatic, and Gaussian. However, due to nonlinear astrophysical processes on small scales that erased primordial information, the properties of primordial perturbations on such scales are completely unknown.\\

The understanding of small scale primordial perturbations is not only crucial to establishing the full picture of inflation but also fascinating as it leads to rich phenomenology that could be observed in the near future, such as scalar-induced gravitational waves (SIGWs) and the formation of primordial black holes (PBHs). Both SIGWs and PBHs are associated with an enhancement of the primordial scalar power spectrum on small scales, while the large scale fluctuations remain nearly scale-invariant. There are numerous inflationary models in which the primordial scalar power spectrum is enhanced by various mechanisms, in single-field models \citep{Leach:2001zf, Byrnes:2018txb, Ozsoy:2019lyy, Dalianis:2018frf, Garcia-Bellido:2017mdw} and in multi-field models \citep{Sasaki:1998ug, Garcia-Bellido:1996mdl, Kawasaki:1997ju, Frampton:2010sw, Giovannini:2010tk, Clesse:2015wea, Inomata:2017okj, Gordon:2000hv, Anguelova:2020nzl, Bhattacharya:2022fze, Braglia:2020eai, Braglia:2020fms, Ema:2020zvg, Cotner:2019ykd, Gundhi:2018wyz, Christodoulidis:2023eiw, Inomata:2018cht, Fumagalli:2020adf, Cai:2021wzd, Meng:2022low, Dimastrogiovanni:2024xvc, Qin:2023lgo, Pi:2017gih, Wang:2024vfv}. 

Let us focus on models of single-field inflation. 
These models can be broadly categorized into two types: those that modify the inflaton potential \citep{Domenech:2023dxx, Pi:2022zxs, Cai:2022erk, Ragavendra:2020sop, Fu:2020lob, Liu:2020oqe, Bhaumik:2019tvl, Mishra:2019pzq, Garcia-Bellido:2016dkw, Yokoyama:1998pt, Ivanov:1994pa, Starobinsky:1992ts, Di:2017ndc, Zhai:2022mpi, Balaji:2022rsy, Choudhury:2024one, Choudhury:2023hfm, Caravano:2024tlp, Cai:2023uhc}, leading to non-trivial evolution of the slow-roll parameters, 
and those that alter the kinetic term or introduce non-minimal couplings to gravity, resulting in modified equations of motion for the scalar perturbations \citep{Karam:2018mft, Fu:2019ttf, Gundhi:2020kzm, Frolovsky:2022ewg, Aldabergenov:2020bpt, Karydas:2021wmx, Canko:2019mud, Chen:2021nio, Kawai:2021edk, Zhang:2021rqs, Yi:2022anu, Lin:2020goi, Yi:2020cut, Yi:2020kmq, Gao:2021vxb, Lin:2021vwc, Ballesteros:2018wlw, Heydari:2021gea, Solbi:2021wbo, Teimoori:2021pte}.

The inflection-point type or ultra-slow-roll (USR) scenario is a widely discussed approach to enhancing the power spectrum by modifying the inflaton potential with a feature that deviates the inflaton from the slow-roll attractor solution. However, the non-slow-roll phase cannot persist for too many e-folds; otherwise, inflation will either end prematurely or continue indefinitely. Therefore, in practice, a healthy and complete model of this type typically involves two phases of standard slow-roll inflation before and after the non-slow-roll phase. The simplest model of this kind is an inflationary model with a piecewise potential, composed of two segments of slow-roll potentials with different parameters. 
It has been extensively studied in single-field models with piecewise linear potentials, which are often called Starobinsky's linear potential model, that the enhancement of curvature perturbations exhibits distinct characteristics, such as dips and growth slopes in the power spectrum, as well as oscillatory behavior \citep{Starobinsky:1992ts, Byrnes:2018txb, Pi:2022zxs, Domenech:2023dxx, Carrilho:2019oqg}. These features indicate that the analysis conducted using separate universe approach (leading order in the gradient expansion) needs further consideration \citep{Leach:2001zf, Jackson:2023obv}.
However, the piecewise linear potential implies that the non-trivial contribution from the second derivative of the potential, $\dd^2 V(\phi)/\dd \phi^2$, has been largely overlooked and remains unexplored in previous studies. On the other hand, it has been shown that a non-vanishing $\dd^2 V(\phi)/\dd \phi^2$ leads to an exponential tail in the probability density function (PDF) of curvature perturbations \citep{Cai:2018dkf, Biagetti:2018pjj, Atal:2019erb, Atal:2019cdz, Pi:2022zxs, Cai:2021zsp, Cai:2022erk}, due to a nonlinear relationship between the comoving curvature $\mathcal{R}$ and the field perturbation on flat slicing $\delta \phi$: $\mathcal{R} \propto -(1/\eta_V)\ln[1+\eta_{V}\delta \phi]$, which could significantly affect PBH formation. Therefore, studying general piecewise potential models with non-zero second derivatives is of great interest and significance.


We thoroughly discuss the primordial power spectrum of curvature perturbations and non-Gaussianity within a general framework of single-field inflation with piecewise quadratic potentials. This paper is organized as follows: In sec.~\ref{sec:background}, we introduce the piecewise quadratic potential and analytically solve for the background evolution in the most general case. Sec.\ref{sec:The enhancement of curvature power spectrum} presents the analytical derivation of the power spectrum of curvature perturbations, along with a discussion of its scaling behavior and IR/UV features. In Sec.\ref{sec: The non-Gaussianity: deltaN calculation}, we examine non-perturbative non-Gaussianity using the $\delta N$ formalism. Finally, in sec.~\ref{sec:PBHs with non-Gaussian tail}, we explore the implications of non-Gaussianity for PBH formation.

\section{The background}
\label{sec:background}
\begin{figure}[htbp]
\centering
\includegraphics[width=.8\textwidth]{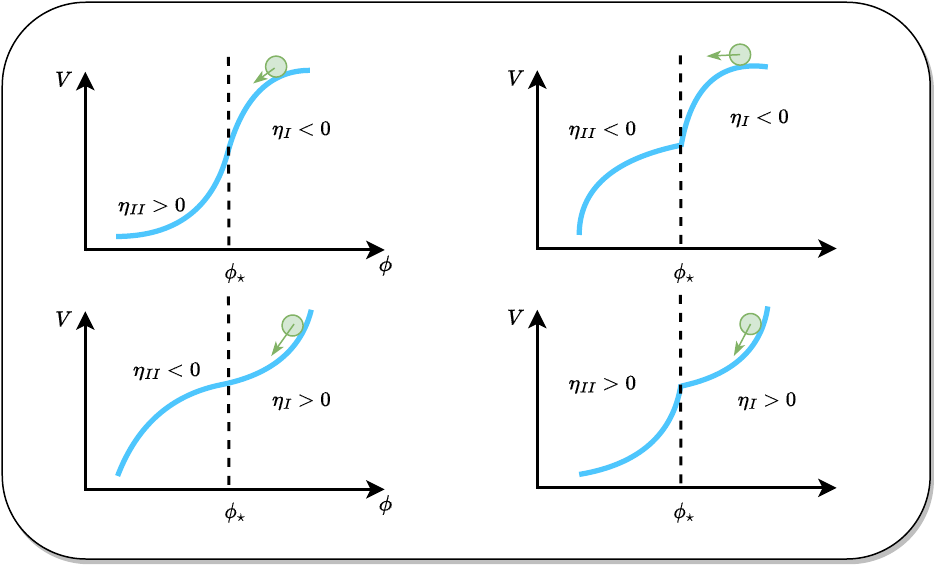}
\caption{A schematic diagram of the potentials we are interested in.\label{fig:1}}
\end{figure}

We parameterize a piecewise potential in the vicinity of its junction point as $V_\star=V(\phi_\star)$ up to the second order in $\phi-\phi_\star$, 
\begin{equation}
    V=\left\{
    \begin{aligned}
          &V_\star\left(1+\sqrt{2\epsilon_{I}}(\phi-\phi_\star)+\frac{1}{2}\eta_{I}(\phi-\phi_\star)^2\right)\quad{\rm for}~ \phi\geq\phi_\star~,\\
          &V_\star\left(1+\sqrt{2\epsilon_{II}}(\phi-\phi_\star)+\frac{1}{2}\eta_{II}(\phi-\phi_\star)^2\right)\quad{\rm for}~ \phi<\phi_\star~.
    \end{aligned}
    \right.
    \label{potential}
\end{equation}
 In the scenario considered here, we do not have an upward step, that is, the case of $\Delta V = 0$ in \citep{Cai:2022erk,Cai:2021zsp}. 
The slow-roll parameters we used in the following discussion are defined by $\epsilon_V\equiv M_{\mathrm{p}}^2(V_{,\phi}/V)^2/2$ and $\eta_V\equiv M_{\mathrm{p}}^{2}V_{,\phi\phi}/V$, where $M_{\mathrm{p}}$ represents the reduced Planck mass. 
In the above, as well as in what follows,
$\epsilon_X$ and $\eta_X$ ($X=I,\,II$) are the potential parameters, respectively, in the regions $I$ ($\phi\geq\phi_\star$) and $II$ ($\phi\leq\phi_\star)$.
As shown in fig.~\ref{fig:1}, we consider four different piecewise potentials.

The Friedmann equations and the scalar field equation on the flat Friedmann-Lemaître-Robertson-Walker (FLRW) background are given by
\begin{align}
    &3M_{\mathrm{p}}^2H^2=\frac{1}{2}\dot\phi^2+V(\phi),\\
    &-2M_{\mathrm{p}}^2\dot H=\dot\phi^2,\\
    &\ddot\phi+3H\dot\phi+V_{,\phi}=0.
\end{align}
For simplicity, we adopt the reduced Planck unis $M_\mathrm{p}=1$ in the remaining part of the paper. 
We introduce the e-folding number $\dd n =H\dd t$ as the time variable to solve the background field equation,
\begin{align}
       &\frac{d^2\phi}{dn^2}+(3-\epsilon_H)\frac{d\phi}{dn}+\frac{V_{,\phi}}{H^2}=0\,,
\end{align}
where $\epsilon_H=-\dot H/H^2$, which is approximately equal to $\epsilon_V$.
Our primary interest lies in the local dynamics of the inflaton around the joint point. Inserting the potential \eqref{potential} into the above and assuming $\epsilon_{H}\ll1 $, we have $H^2=V_{\star}/3$, and the equation for $\phi$ may be written as
\begin{align}\label{eq:Background equation in delta N}
    \frac{d^2\widetilde{\phi}_{X}}{dn^2}+3  \frac{d\widetilde{\phi}_{X}}{dn}+3\eta_{X}\widetilde\phi_X=0.
\end{align}
where $X =I,\,II$, and $\widetilde{\phi}_X \equiv \phi - \phi_\star + \sqrt{2 \epsilon_{X}}/\eta_{X}$. 
Requiring the monotonicity of $\phi(n)$, we assume $\eta_{X} < 3/4$.
Then, the general solution for $\phi$ is given by
\begin{equation}
\begin{aligned} 
   {\phi}_X &= C_{X,-}e^{-\lambda_{X,-}(n-n_\star)} + C_{X,+}e^{-\lambda_{X,+}(n-n_\star)}+\phi_{X,0}\,;\quad
   \phi_{X,0}=\phi_\star-\frac{\sqrt{2\epsilon_{X}}}{\eta_{X}}~,
    \label{eq:phi solution}
\end{aligned}
\end{equation}
where $-\lambda_{X,\pm}$ are the characteristic roots of Eq. \eqref{eq:Background equation in delta N}, given by
\begin{align}
    \lambda_{X,\pm} =\pm \nu_{X}+\frac{3}{2}\,;\quad
    \nu_{X}\equiv\sqrt{\frac{9}{4}-3\eta_X}\,.
\label{lambdadef}
\end{align}
Sometimes it is convenient to express $\eta_X$ in terms of $\nu_X$,
\begin{equation}
    \eta_{X}=\frac{1}{3}\left(\frac{3}{2}-\nu_X\right)\left(\frac{3}{2}+\nu_X\right)=\frac{1}{3}\lambda_{X+}\lambda_{X-}.
\label{etanu}
\end{equation}

\begin{figure}[htbp]
\centering
\includegraphics[width=0.8\textwidth]{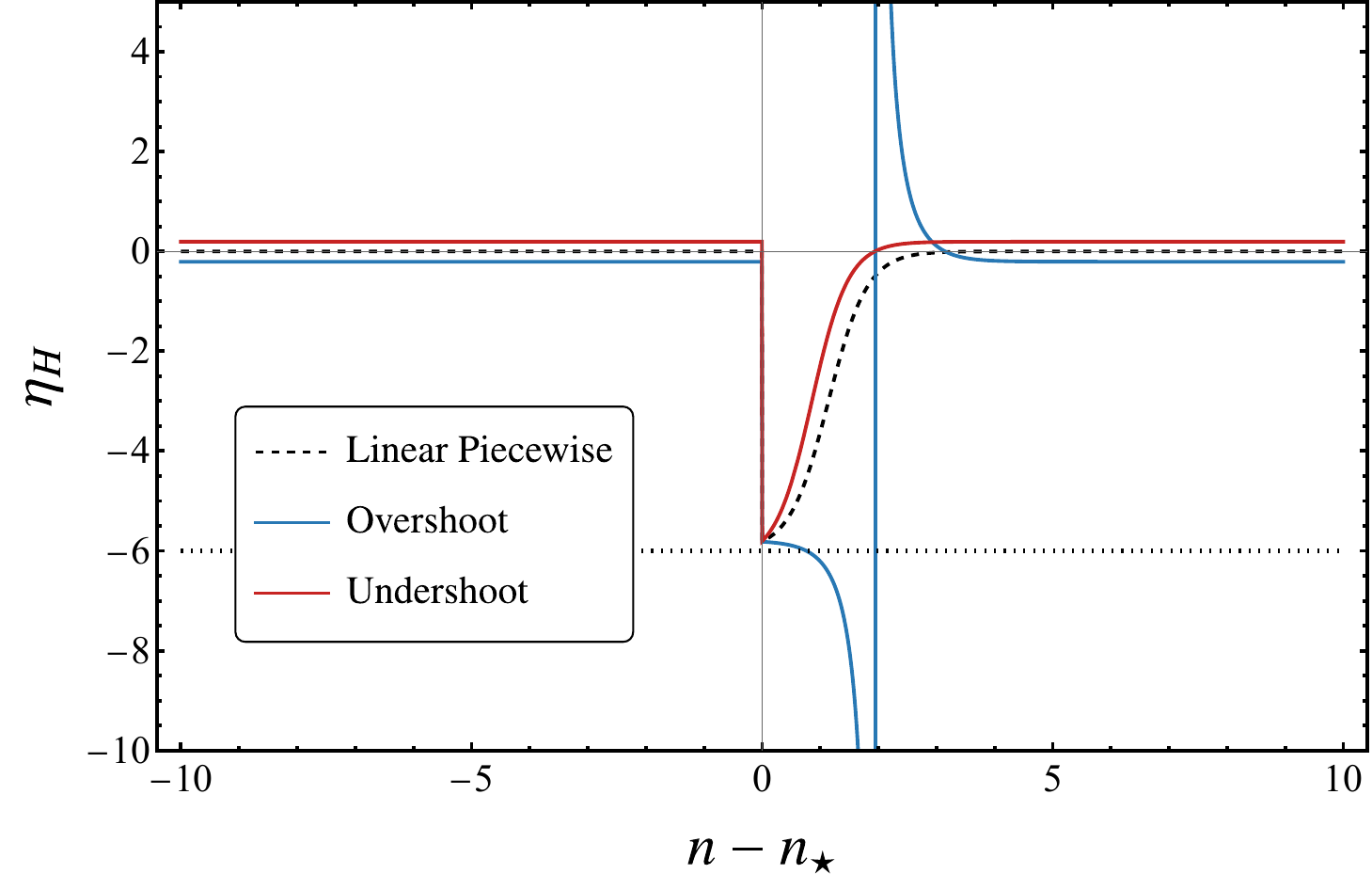}
\caption{\textbf{The time evolution of  $\eta_{H}$ for different cases}. $\eta_{I}=\eta_{II}=0.1,\eta_{I}=\eta_{II}=-0.1$ are used for the blue curve and red curve separately. For the overshoot case which is shown by the blue curve, corresponding to $\alpha>1$, $\eta_{H}$ diverges because $\dot\phi$ hits 0 before the inflaton settles down to the attractor phase. The undershoot case is shown by the red curve, corresponding to $\alpha<1$, which indicates that the inflaton enters the second attractor phase before it hits the potential minimum. }
\label{etaplot}
\end{figure}

Given the initial condition $(\phi_{i},\pi_{i} \equiv(\dd\phi/\dd n)_{i})$ , the coefficients for the two linearly independent solutions can be determined as
\begin{equation}
\begin{aligned}
        C_{X,\pm}
        = \pm \frac{\pi_\star +\lambda_{X,\mp} \widetilde{\phi}_{X,\star}}{\lambda_{X,-} - \lambda_{X,+}},
        \label{backgroundcoe}
    \end{aligned}
    \end{equation}
for the piecewise potential given by \eqref{potential}, with $\eta_{X}<3/4$, 
$-\lambda_{X,+}=-\lambda_{X,-}-2\mu_X<-\lambda_{X,-}$. 
Hence the solution with the characteristic exponent $-\lambda_{X,+}$ always decays faster than the other solution with $-\lambda_{X,-}$. 
The latter solution corresponds to the solution in the slow-roll limit, the so-called attractor solution, while we call the solution with $-\lambda_{X,+}$ the decaying solution.
We focus on the initial condition for which the decaying solution has already disappeared and the solution is in the attractor regime before $\phi$ reaches $\phi_\star$. 
Namely, we assume $C_{I,+}=0$. This gives $C_{I,-}=\sqrt{2\epsilon_I}/\eta_I$.
Under this assumption, when $\phi$ reaches $\phi_\star$, we have 
\begin{equation}
\pi_{\star,I} = -\lambda_{I,-}\frac{\sqrt{2\epsilon_{I}}}{\eta_{I}}\,.
\label{pistar}
\end{equation}
Since $\pi$ is continuous at $\phi_\star$, $\pi_{\star,I}=\pi_{\star,II}=\pi_\star$. 
This fixes the coefficients $C_{II,\pm}$,
\begin{align}
\label{matchbg}
&C_{II,\pm}(\phi_{\star},\pi_{\star})=\pm \frac{3}{\nu_{II}}\left(\frac{\sqrt{2\epsilon_{I}}}{2\nu_{I}+3}\mp\frac{\sqrt{2\epsilon_{II}}}{2\nu_{II}\pm 3}\right),
\end{align}
which determines the dynamic at the second stage.
When $|C_{X,+}|>|C_{X,-}|$, the decay mode dominates soon after the transition, and thus leads to a period of non-attractor phase (See in fig.~\ref{etaplot}). 
If the background dynamics can be described by only the decaying solution, the second slow-roll parameter reads
\begin{align}
    \eta_{H}\equiv\frac{\dot\epsilon_{H}}{H\epsilon_{H}}=-3-2\nu_{II}\,,
\end{align}
its value becomes $\eta_{H}\approx-6$ if $\eta_{II}\ll 1$, in which case the phase is called ultra-slow-roll (USR) inflation.

\begin{figure}[htbp]
\centering
\includegraphics[width=0.45\textwidth]{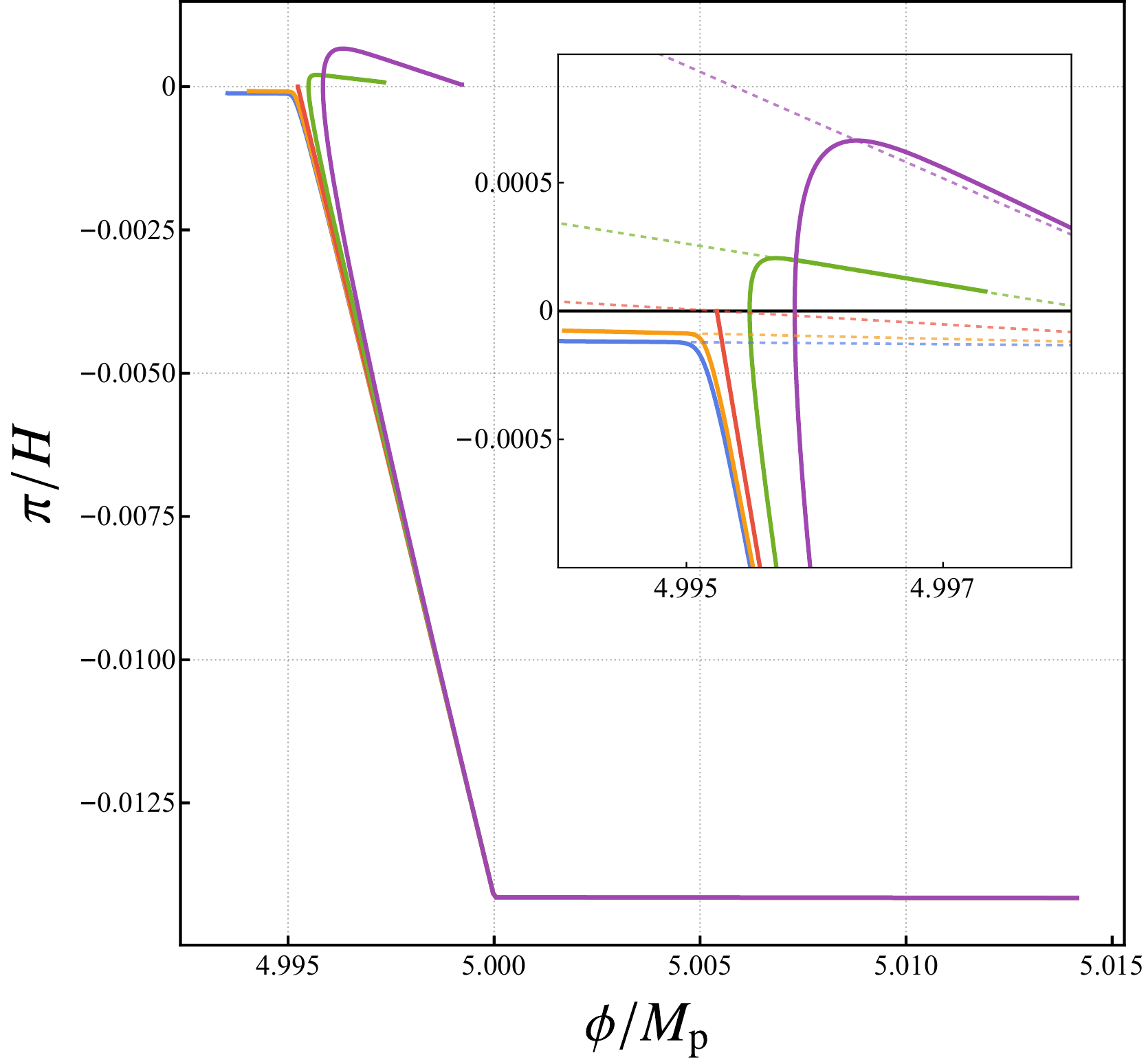}
\includegraphics[width=0.45\textwidth]{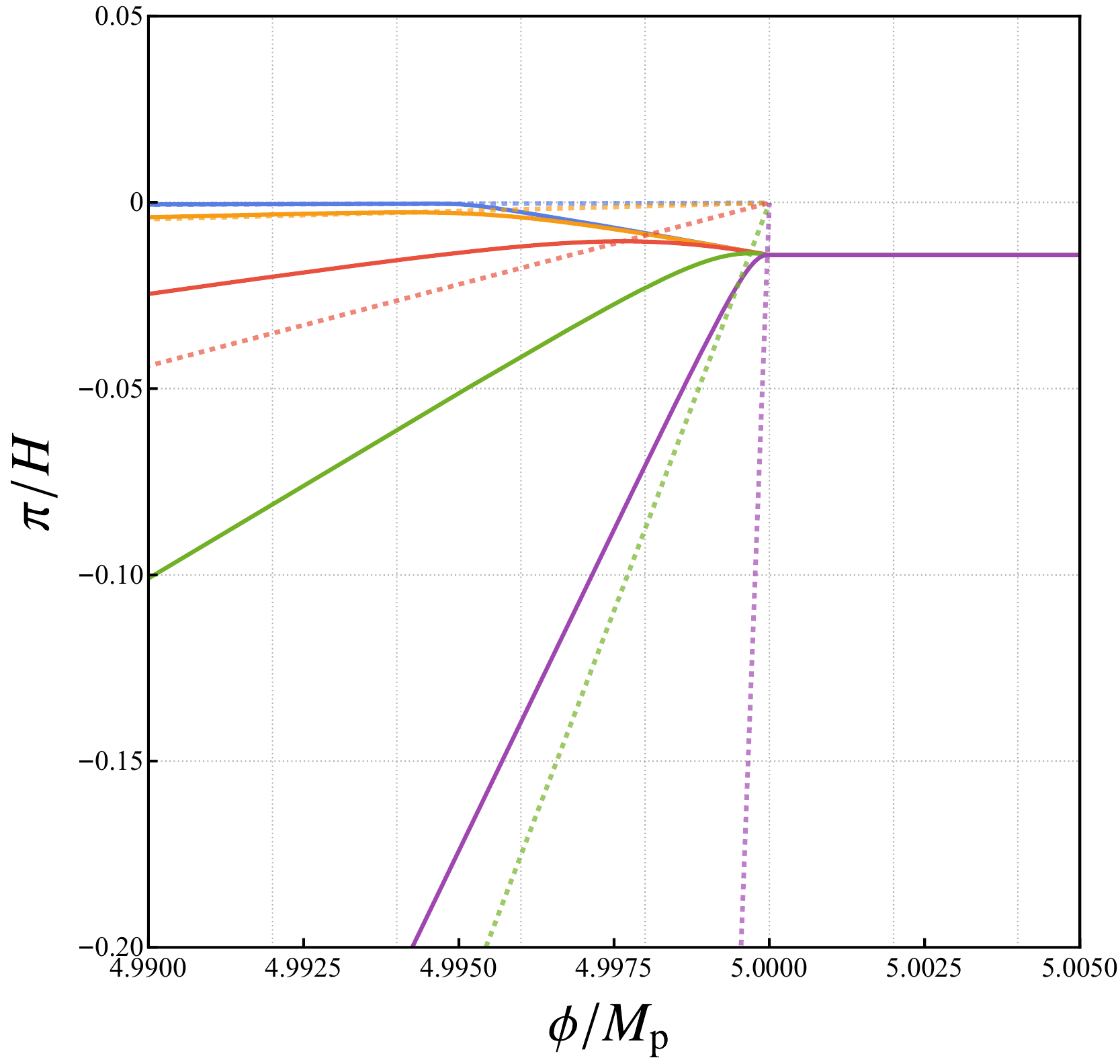}
\caption{\textbf{The background phase diagram} for a positive $\eta_{II}$ (left panel) and negative $\eta_{II}$ (right panel) starting from a slow roll attractor initial condition given during the first stage of inflation. We take $R_{\epsilon}=10^{-2},\eta_{I}=10^{-3}$ and $\phi_{\star}=5M_{\rm p}$ in both figures, the solid lines are given by the numerical solution of the background equations, the dashed lines are the accordingly slow roll attractor solution of the second stage given analytically. In the left panel, we take $\eta_{II}=\eta_{II}^{r+}\times(e^{-2},e^{-1},1,e^{1},e^{2})$ for the blue, orange, red, green and purple lines. In the right panel we take $\eta_{II}=\eta_{II}^{r-}\times(10^{-2},10^{-1},1,10^{1},10^{2})$ for the blue, orange, red, green and purple lines. We use $\eta_{II}^{r+}\approx0.02969$ which is given by $\alpha=0$ for the vanishing non-decay mode case, $\eta_{II}^{r-}=-4.3757$ which is given by $\xi_{2}=0$ defined in \eqref{prcoe} for the vanishing second-order gradient expansion contribution to the curvature perturbation, which will be discussed in the next section.}
\label{background}
\end{figure}

Typically, the relaxation period is short. However, re-writing \eqref{matchbg} as 
\begin{align}
   C_{II,-}=\frac{3}{\nu_{II}}\frac{\sqrt{2\epsilon_{II}}}{2\nu_{II}-3}(\alpha-1)\,,
   \label{CII+}
\end{align}
where $\alpha$ is defined as
\begin{align}
    \alpha\equiv\frac{3-2\nu_{II}}{(2\nu_{I}+3){R_\epsilon}}=\frac{\lambda_{II,-}}{\lambda_{I,+}R_\epsilon}\,;
    \quad R_\epsilon=\sqrt{\epsilon_{II}/\epsilon_{I}}\,,
    \label{alpha}
\end{align}
we see that $C_{II,-}$ vanishes for the parameters that satisfy $\alpha=1$.
In this case, the relaxation stage becomes infinitely long.
It is worth mentioning is that the magnitude of $\alpha$ depends linearly on $\eta_{II}$ in the limit $\eta_{II}\ll1$, 
\begin{align}
    \alpha\approx\frac{\eta_{II}}{R_\epsilon(\nu_I+3/2)}\,\quad {\rm for}~|\eta_{II}|\ll1
    \label{approxalpha}
\end{align}
Thus $\alpha=1$ may be realized for the value of $\eta_{II}$ given by $\eta_{II}\approx(\nu_I+3/2)R_\epsilon$ when $R_\epsilon\ll1$.
During the infinitely long period of approaching the asymptotic solution, the inflaton stays in a non-slow roll state (see in fig.~\ref{background}, the red line in the top panel). 

We note that this eternal USR-like case is extremely unstable, that even a tiny deviation of the boundary condition at the joint point $(\phi_\star,\dot\phi_\star)$ or a slight time evolution of $H$ can violate this situation, i.e. introduce a non-vanishing attractor mode that eventually
dominates the background solution. 
Thus, in reality the parameters that satisfy $\alpha=1$ can only lead to a relatively long USR-like phase.
Nevertheless, it is of interest if we can tune the potential in such a way that it exactly leads to an eternal USR-like stage. We leave this issue for future study.

The parameter $\alpha$ is also useful when considering the asymptotic behavior of $\phi$.
The sign of $C_{II,-}$ determines the sign of the asymptotic value of the momentum $\pi$, together with the sign of $\eta_{II}$, as the sign of $\pi(=d\phi/dn)$ is determined by that of the product $\lambda_{II,-}C_{II,-}\propto (\alpha-1)$.
If $\eta_{II}<0$, $\alpha<0$. Hence the inflaton approaches the attractor solution smoothly.
On the other hand, if $\eta_{II}>0$, two kinds of behavior appear depending on the value of $\alpha$.
When 1) $\alpha<1$, the attractor phase is realized 
before crossing the minimum of the potential $\phi_{X,0}$
(the blue and orange lines in the top panel of fig.~\ref{background}),
while when 2) $\alpha>1$, the attractor phase is realized 
after crossing $\phi_{X,0}$ (the purple and green lines in the top panel of fig.~\ref{background}).
Namely, it enters the attractor phase after the moment it starts to roll back. 

\begin{figure}[htbp]
\centering
\includegraphics[width=1\textwidth]{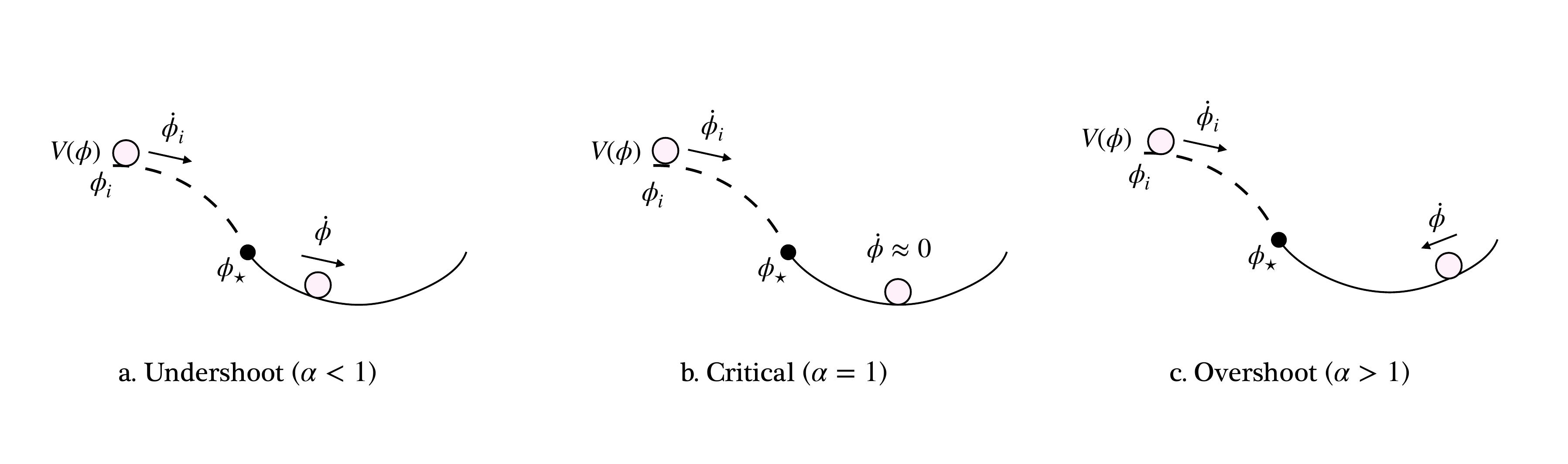}
\caption{\textbf{Schematic plots for the potential and  inflaton evolution when $\eta_{II}>0$}. In each of the plots, The left point shows the initial attractor state while the right point shows the final attractor state of the inflaton. Plot (a) shows the undershoot case that inflaton goes back to the attractor phase before reaching the potential minimum. Plot (c) shows the overshoot case that the inflaton goes back to the attractor phase after when it passes the potential minimum and makes a U-turn. Plot (b) shows the critical case for the eternal USR-like phase case, that the inflaton stops by the potential minimum.}
\end{figure}

Here, it is worth mentioning that we assume that the end of inflation is in the attractor phase. If the attractor phase is realized after a U-turn, we may need a special mechanism to end inflation. We simply assume the existence of such a mechanism without discussing any specific models in this paper.
If we would relax this attractor assumption, we would have to keep track of the time evolution of the curvature perturbation to determine its final spectrum, instead of taking the asymptotic future limit. As this would complicate our discussion significantly, we leave studies of such cases for the future.

In order to have some intuition about the power spectrum enhancement, we evaluate the ratio of power for the small scale modes that exit the horizon after the inflaton enters the attractor phase during the second stage to the large scale modes that exit the horizon during the first stage. Assuming small $\eta$ limit, the power spectrum is nearly scale-invariant on both the long and short wavelength ends, and the ratio is thus given by
\begin{align}
    \frac{\mathcal{P}_{\mathcal R}(k_{UV})}{\mathcal{P}_{\mathcal R}(k_{IR})}\approx\frac{\left.H^2/\epsilon_{H}\right|_{t=t_\star}}{\left.H^2/\epsilon_{H}\right|_{t\gg t_\star}}&=\frac{C_{I,-}^2\lambda_{I,-}^2}{C_{II,-}^2\lambda_{II,-}^2}
    \nonumber\\
    &
    =\left(\frac{\lambda_{II,-}-\lambda_{II,+}}{\lambda_{II,-}-\lambda_{I,+}R_{\epsilon}}\right)^2
    \nonumber\\
    &=\frac{1}{(1-\alpha)^2}\left(\frac{4\nu_{II}}{(3+2\nu_{I})R_{\epsilon}}\right)^2,
    \label{prestim}
\end{align}
which is a good approximation to the order of enhancement of the power spectrum due to the transition regardless of a small spectral index induced by nonzero $\eta$ and $\epsilon$. The upper expression easily degenerates to the standard slow roll version $\mathcal{P}_{\mathcal R }(k_{UV})/{\mathcal{P}_{\mathcal R}(k_{IR})}=R_\epsilon$ when taking $\nu_{II}=\nu_{I}=3/2$ ~(or $\eta_{I},\eta_{II}=0$, or equivalently $\lambda_{I,-}=\lambda_{II,-}=0,\lambda_{I,+}=\lambda_{II,+}=3$ ).  However, $\alpha$ starts to determine the enhancement in cases that $R_{\epsilon}\neq1$ (see an example in the next section fig.~\ref{uturn}).

\section{The curvature power spectrum}\label{sec:The enhancement of curvature power spectrum}
\subsection{The perturbation equations}
We consider the metric with scalar-type linear perturbations, together with the scalar field perturbation,
\begin{align}
    &ds^2=-(1+2A)dt^2+2aB_{,i}dx^idt+a^2[1+2\calR\delta_{ij}+2E_{,ij}]dx^idx^j,\\
    &\phi(\mathbf{x},t)=\phi(t)+\delta\phi(\mathbf{x},t)
    \end{align}
Among various choices of gauges, the one most convenient for evaluating the quantum vacuum fluctuations is the flat slicing where we set $\calR=0$. Another is the slicing on which the fluctuation in the scalar field vanishes $\delta\phi=0$, called the comoving slicing. The field fluctuation on flat slicing, denoted by $\delta\phi_f$ and the curvature perturbation on comoving slicing, denoted by $\calR$ are related by
\begin{align}
    \calR&=-\frac{H}{\dot\phi}\delta\phi_f.
\label{calRdelphi}
\end{align}
On flat slicing, the equation of motion for the Fourier mode of $\delta\phi_f$, denoted by $\delta\phi_k$, is given by
\begin{align}
\ddot{\delta\phi_k}+3H\dot{\delta\phi_k}+\left(\frac{k^2}{a^2}+V_{,\phi\phi}\right)\delta\phi_k=\frac{1}{a^3}\frac{d}{dt}\left(\frac{a^3}{H}\dot\phi^2\right)\delta\phi_k\,.
\label{dfeom}
\end{align}
By introducing the conformal time $d\tau=dt/a$, the corresponding equation for the Fourier mode of $\calR_c$, denoted by $\calR_k$, becomes 
\begin{align}
\mathcal{R}_k''+2\frac{z'}{z}\mathcal{R}_k'+k^2\mathcal{R}_k=0,
    \label{reom}
\end{align}
where $z\equiv a d\phi/dn$. Under the slow roll approximation $|\epsilon|\ll 1$, the dimensionless conformal time derivatives of $z$ are given
\begin{align}
    &\frac{z'}{z}=-\tau^{-1}\left(2\frac{d\phi}{dn}+\frac{V_{,\phi}}{H^2}\right),\\
    &\frac{z''}{z}=\tau^{-2}\left(2-\frac{V_{,\phi\phi}}{H^2}\right).
    \label{zdp}
\end{align}
It is worth mentioning that the first derivative of
$z$ depends explicitly on the background dynamics, while the second derivative does not.

In our model, $z$ is a continuous function of the scale factor $a$,
\begin{align}
    z\simeq\left\{
    \begin{aligned}
        &a_\star\left(\frac{a}{a_\star}\right)^{\nu_{I}-\frac{1}{2}}\frac{3\sqrt{2\epsilon_{I}}}{\nu_{I}+3/2}\quad{\rm for}~\phi>\phi_\star,\\
        &a_\star\left(\frac{a}{a_\star}\right)^{\nu_{II}-\frac{1}{2}}\frac{3{\sqrt{2\epsilon_{II}}}}{2\nu_{II}}\left(1+\frac{2\nu_{II}-3}{(2\nu_{I}+3)R_{\epsilon}}\right)
       \\&-a_\star\left(\frac{a}{a_\star}\right)^{-\nu_{II}-\frac{1}{2}}\frac{3\sqrt{2\epsilon_{II}}}{2\nu_{II}}\left(1-\frac{2\nu_{II}+3}{(2\nu_{I}+3)R_{\epsilon}}\right)\quad{\rm for}~\phi<\phi_\star.
    \end{aligned}\right.
\end{align}
Yet $z'/z$ appears to have a discontinuity at $a_\star$. 
By taking the limit $\epsilon_H\ll1$ and setting $\tau\approx -1/(aH_\star)$,
the discontinuity of $z'/z$ at $\phi_\star$ is found as
\begin{align}
    \frac{z'}{z}\simeq\left\{
    \begin{aligned}
        &\left(\frac{1}{2}-\nu_I\right)\tau_\star^{-1}\quad\phi-\phi_\star=0^{+},\\
        &\left[2-R_{\epsilon}\left(\frac{3}{2}+\nu_I\right)\right]\tau_\star^{-1}\quad\phi-\phi_\star=0^-.
    \end{aligned}\right.
\end{align}
As a result, the second derivative of $\mathcal{R}_k$ appears to be discontinuous at the joint point. 
To circumvent this potential issue/complexity, we introduce a new variable,
\begin{align}
    u=a\delta\phi=z\mathcal{R},
\end{align}
and rewrite \eqref{reom} as
\begin{align}
     u_{k}''+\left(k^2 - \frac{z''}{z}\right)u_{k}=0\,.
     \label{ms}
\end{align}
A convenient property of this form of the equation is that, as \eqref{zdp} tells us, $z''/z$ depends only on the potential parameter,
\begin{equation}
\begin{aligned}
    \frac{z''}{z}&\simeq\left\{\begin{aligned}
        &\tau^{-2}(2-3\eta_{I})\quad{\rm for}~\phi>\phi_\star,\\
        &\tau^{-2}(2-3\eta_{II})\quad{\rm for}~\phi<\phi_\star,
    \end{aligned}
    \right.
    \end{aligned}
\end{equation}
where $\epsilon_H\ll1$ is assumed. Hence under this approximation, the general solution to \eqref{ms} is analytically found as
\begin{align}
    u_{k,I}=C_{I}\sqrt{-k\tau}H^{(1)}_{\nu_I}(-k\tau)+D_{I}\sqrt{-k\tau}H^{(2)}_{\nu_I}(-k\tau)\quad{\rm for}~\phi>\phi_\star\,,\\
    u_{k,II}=C_{II}\sqrt{-k\tau}H^{(1)}_{\nu_{II}}(-k\tau)+D_{II}\sqrt{-k\tau}H^{(2)}_{\nu_{II}}(-k\tau)\quad{\rm for}~ \phi<\phi_\star\,.
    \label{u2}
\end{align}
where $C_X$ and $D_X$ ($X=I,\,II$) are constants, and $H_\nu^{(1)}(x)$ and $H_{\nu}^{(2)}(x)$ are the Hankel functions of the first kind and second kind, respectively.
In the limit of $-k\tau\rightarrow\infty$, the $k$-mode is deep inside the horizon. 
The Hankel function in this limit is
\begin{align}
    H_{\nu}^{(1)}(x\rightarrow\infty)\rightarrow\sqrt{\frac{2}{\pi x}}\exp\left({ix-\frac{i\pi}{4}(2\nu+1)}\right),
\end{align}
Thus we obtain the coefficients for the adiabatic vacuum as
\begin{align}
    C_{I}(k)=\frac{\sqrt\pi}{2\sqrt{k}}\exp\left[\frac{i\pi}{4}(2\nu+1)\right],\quad D_I=0\,,
    \label{adiabaticvac}
\end{align}
where the phase is chosen for convenience. Note that $\nu$ is real in our model.

\subsection{Matching at the joint point}
Since $\mathcal{R}_k=u_k/z$ and its time derivative is continuous across the transition, we match  $\mathcal{R}_k$ and  $\mathcal{R}'_k$ at the junction point,
\begin{align}
    \frac{u_{\star,I}}{z_I}&=\frac{u_{\star,II}}{z_{II}}\\
    \frac{u'_{\star,I}}{z_I}-\frac{u_{\star,I}}{z_I}\frac{z_I'}{z_I}&=\frac{u'_{\star,II}}{z_{II}}-\frac{u_{\star,{II}}}{z_{II}}\frac{z_{II}'}{z_{II}}\,.
\end{align}
For the sake of generality, we keep $D_I$ as if it were non-vanishing, which enables us to obtain the power spectrum for a non-adiabatic vacuum initial condition. 
Denoting $x=-k\eta_\star$, the coefficients are given by
\begin{equation}
    \begin{aligned}
    &C_{II}-D_{II}=\\&-\frac{i\pi}{4}(C_{I}+D_{I})\left(2xJ_{\nu_I^-}(x)J_{\nu_{II}}(x)-2xJ_{\nu_{II}^-}(x)J_{\nu_{I}}(x)+(\lambda_{II}^{+}-R_{\epsilon}\lambda_{I}^{+})J_{\nu_{II}}(x)J_{\nu_{I}}(x)\right)\\
    &-\frac{i\pi}{4}(C_{I}-D_{I})\left(2xJ_{\nu_{II}^-}(x)Y_{\nu_{I}}(x)-2xJ_{\nu_{II}}(x)Y_{\nu_I^-}(x)+(\lambda_{II}^{+}-R_{\epsilon}\lambda_{I}^{+})J_{\nu_{II}}(x)Y_{\nu_{I}}(x)\right)\,,
    \label{cdana}
\end{aligned}
\end{equation}
where $J_{\nu}$ and $Y_{\nu}$ are the Bessel functions of order $\nu$, and we have introduced
\begin{equation}
\begin{aligned}
     &\nu_{I}^{\pm}=\pm1+\nu_I,\\
    &\nu_{II}^{\pm}=\pm1+\nu_{II},\\
\end{aligned}
\end{equation}
for notational simplicity.

Now let us evaluate the final amplitude of the curvature perturbation, assuming that the end of the second stage is the end of inflation. We consider the late stage when $\tau\gg\tau_{\star}$ so that all the modes of interest are well outside the horizon, $-k\tau=(a_kH_k/aH)\rightarrow0$. In this limit, we may use the asymptotic behavior of $H_{\nu}^{(1,2)}(x)\rightarrow\pm\Gamma(\nu)(2/x)^\nu/(i\pi)$ to simplify \eqref{u2},
\begin{align}
     &u_{k,II}(-k\tau\rightarrow0)=\sqrt{-k\tau}\Gamma(\nu_{II})\left(-\frac{2}{k\tau}\right)^{\nu_{II}}\frac{1}{i\pi}\left(C_{II}-D_{II}\right).
\end{align}
Then, the curvature perturbation at the end of inflation is obtained as
\begin{align}
   &\mathcal{P}_\mathcal{R}(k)= \frac{k^{3}}{2\pi^2}\left|\mathcal{R}_k\right|^2=\frac{k^{3}}{2\pi^2}\left|\frac{u_{k,{II}}}{z_{II}}\right|^2
    \nonumber\\
   &\quad \overset{-k\tau\rightarrow0}{\longrightarrow}\left(\frac{H_\star}{2\pi}\right)^2\left(\frac{k/k_\star}{2}\right)^{3-2\nu_{II}}\left(\frac{\Gamma(\nu_{II})}{\sqrt{\pi}/2}\right)^2\left|\frac{C_{II}-D_{II}}{\sqrt{\pi/k}/2}\right|^2\left(\frac{2\nu_{II}}{3(1-\alpha)\sqrt{2\epsilon_{II}}}\right)^2\,.
    \label{analyticalpowerspectrum}
\end{align}
In the limit of no transition ($R_{\epsilon}=1, \eta_{II}=\eta_{I}$), the above expression reduces to
\begin{align}
      \mathcal{P}_{\mathcal{R}}(k)=\left(\frac{H_\star}{2\pi} \right)^2\left(\frac{k/k_\star}{2}\right)^{3-2\nu_{I}}\left(\frac{\Gamma(\nu_{I})}{\sqrt{\pi}/2}\frac{2\nu_{I}+3}{6\sqrt{2\epsilon_{I}}}\right)^2\overset{\eta_I\ll1}{\longrightarrow}\frac{H^2}{8\pi^2\epsilon_{I}},
\end{align}
which agrees with the standard slow-roll result. 
For ease in calculating the formalism $\delta N$ after the background solution converges to the attractor phase, the asymptotic expression of $\delta\phi_k$ is given 
\begin{align}
    \mathcal{P}_\mathcal{\delta\phi}(k)=\frac{k^{3}}{2\pi^2}\left|\frac{u_{k,II}(-k\tau\rightarrow0)}{a}\right|^2=\left(\frac{H}{2\pi}\right)^2\left(\frac{k/(aH)}{2}\right)^{3-2\nu_{II}}\left(\frac{\Gamma(\nu_{II})}{\sqrt{\pi}/2}\right)^2\left|\frac{C_{II}-D_{II}}{\sqrt{\pi/k}/2}\right|^2.
    \label{deltaphi}
\end{align}


\begin{figure}[htbp]
\centering
\includegraphics[width=1.\textwidth]{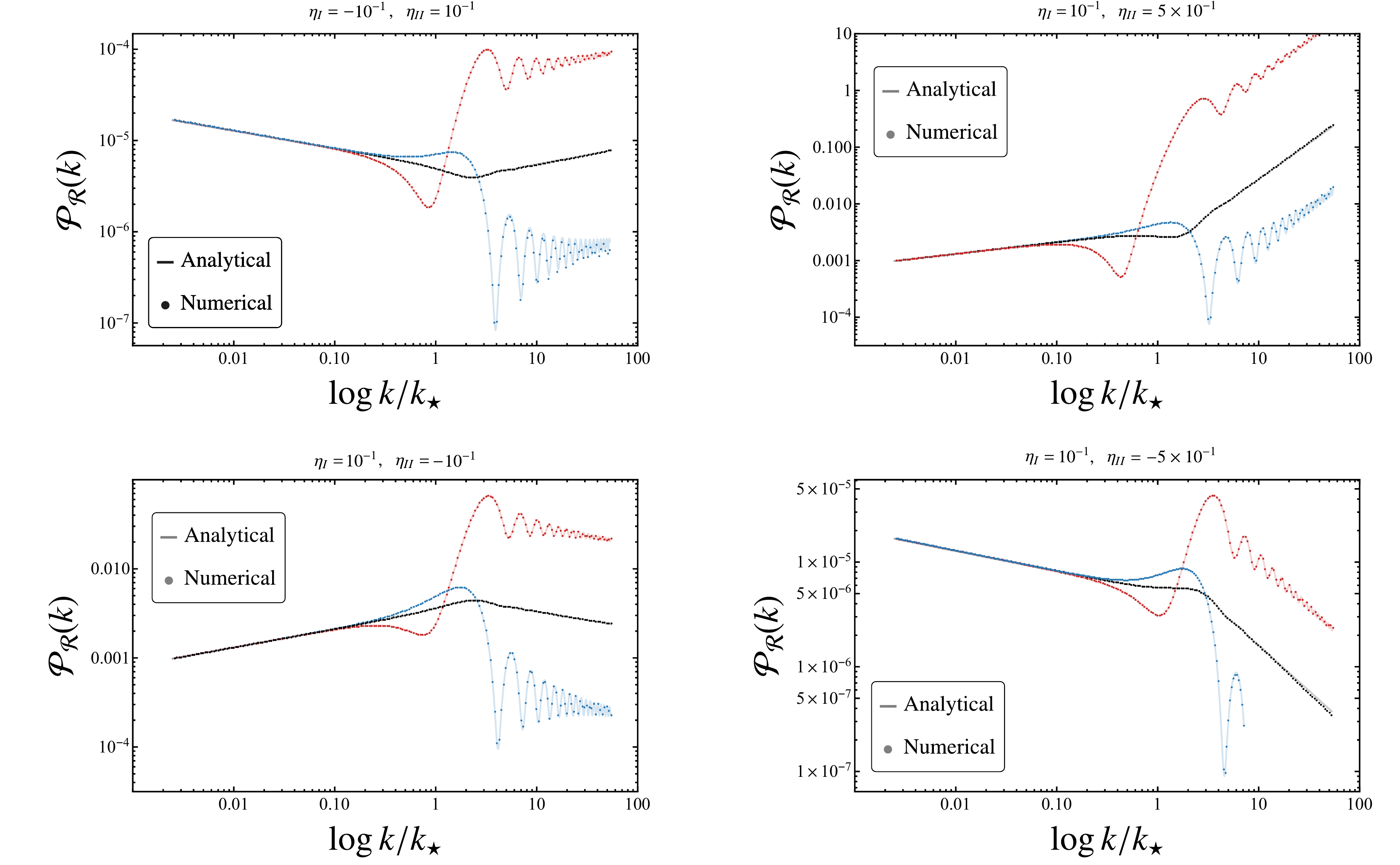}
\caption{\textbf{Primordial power spectrum for the piecewise quadratic model.}
The positions of panels are arranged in accordance with those in fig.~ \ref{fig:1}, classified by the signs of $\eta_{I}$ and $\eta_{II}$. 
The values of $R_\epsilon$ are $R_\epsilon^2=0.1$ (black), $0$ (red), and $10$ (blue).  }
\label{anavsnumresult}
\end{figure}

\begin{figure}[htbp]
\centering
\includegraphics[width=0.45\textwidth]{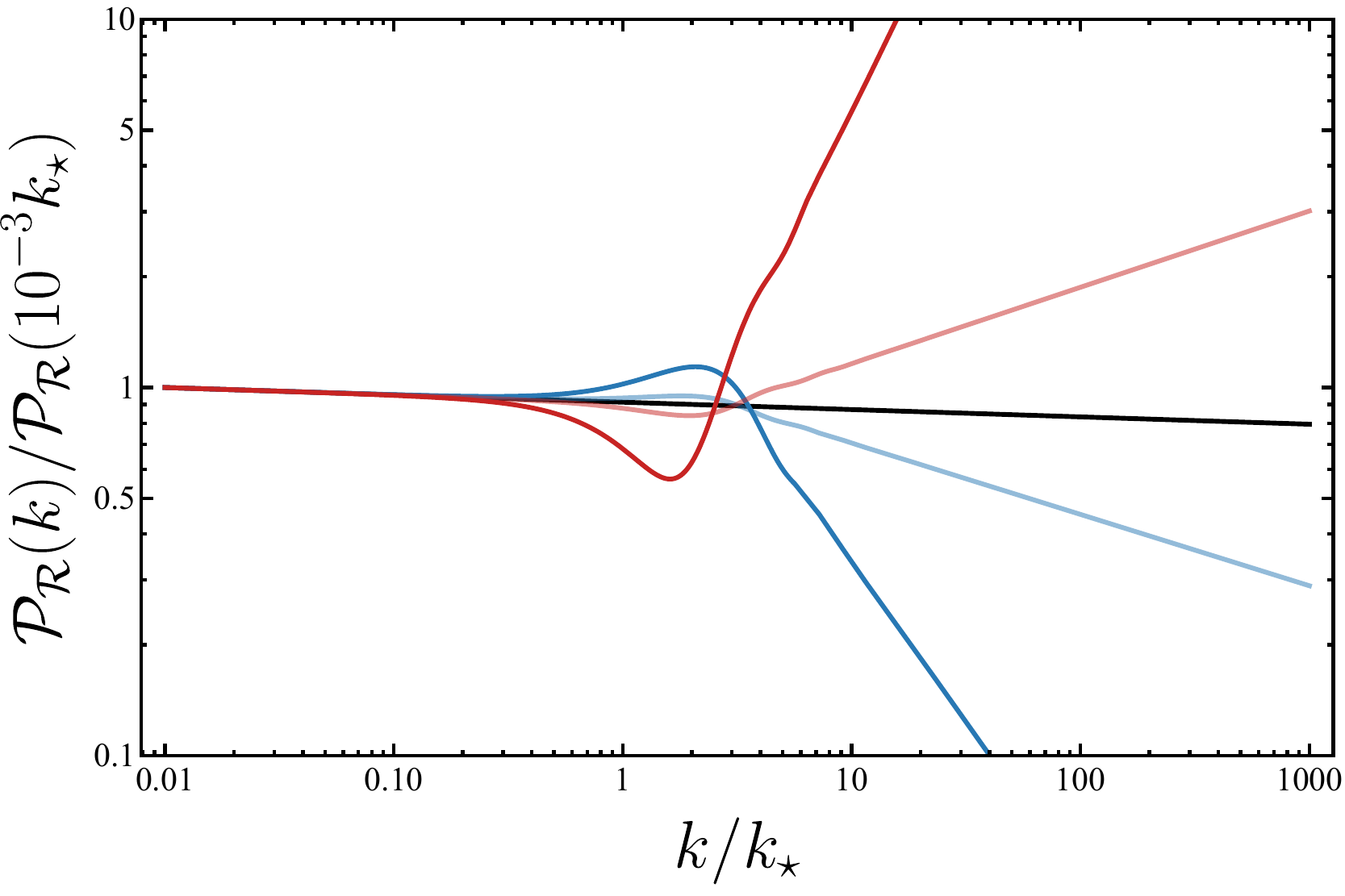}
\hfill\includegraphics[width=0.45\textwidth]{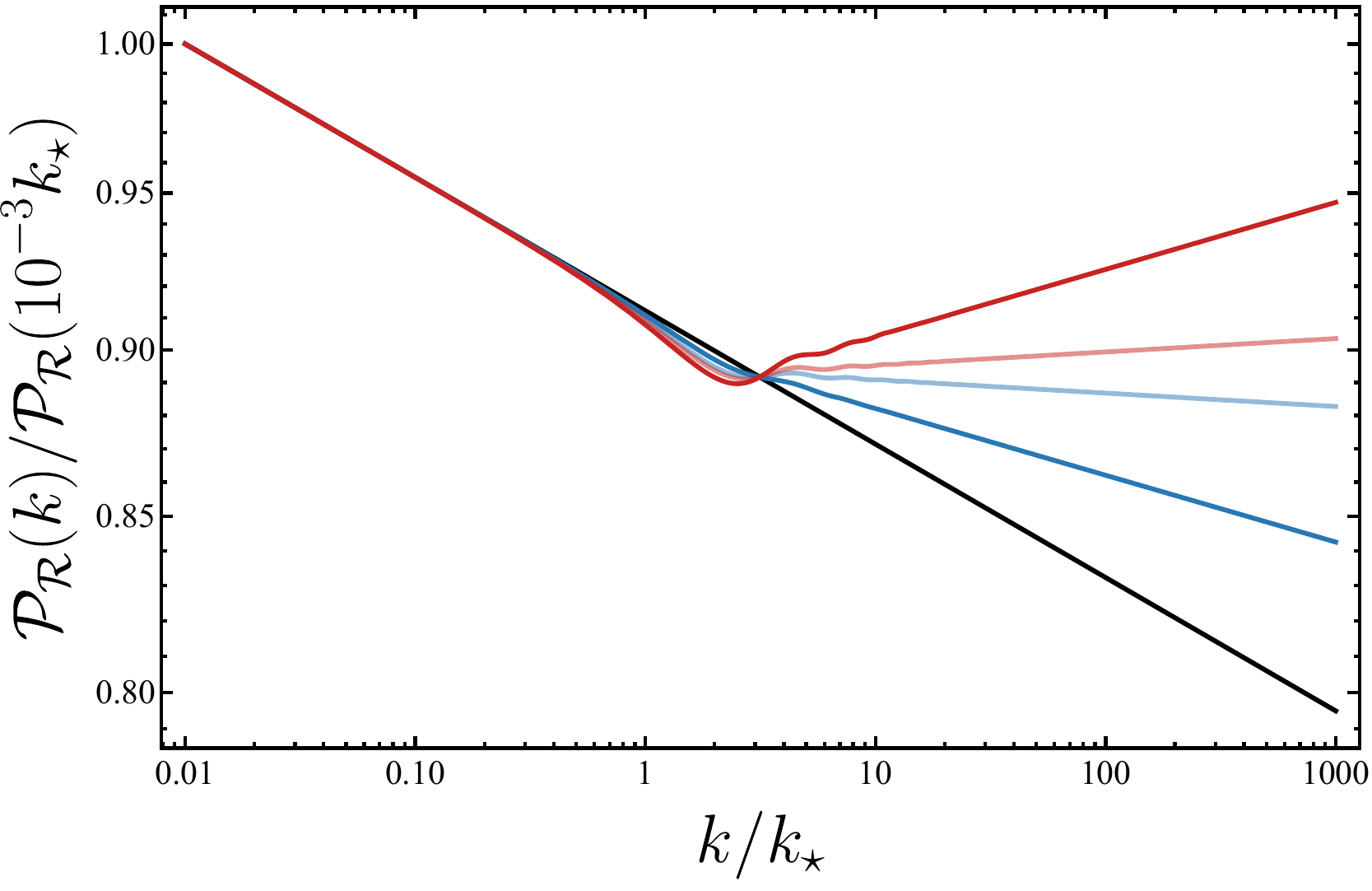}
\includegraphics[width=0.45\textwidth]{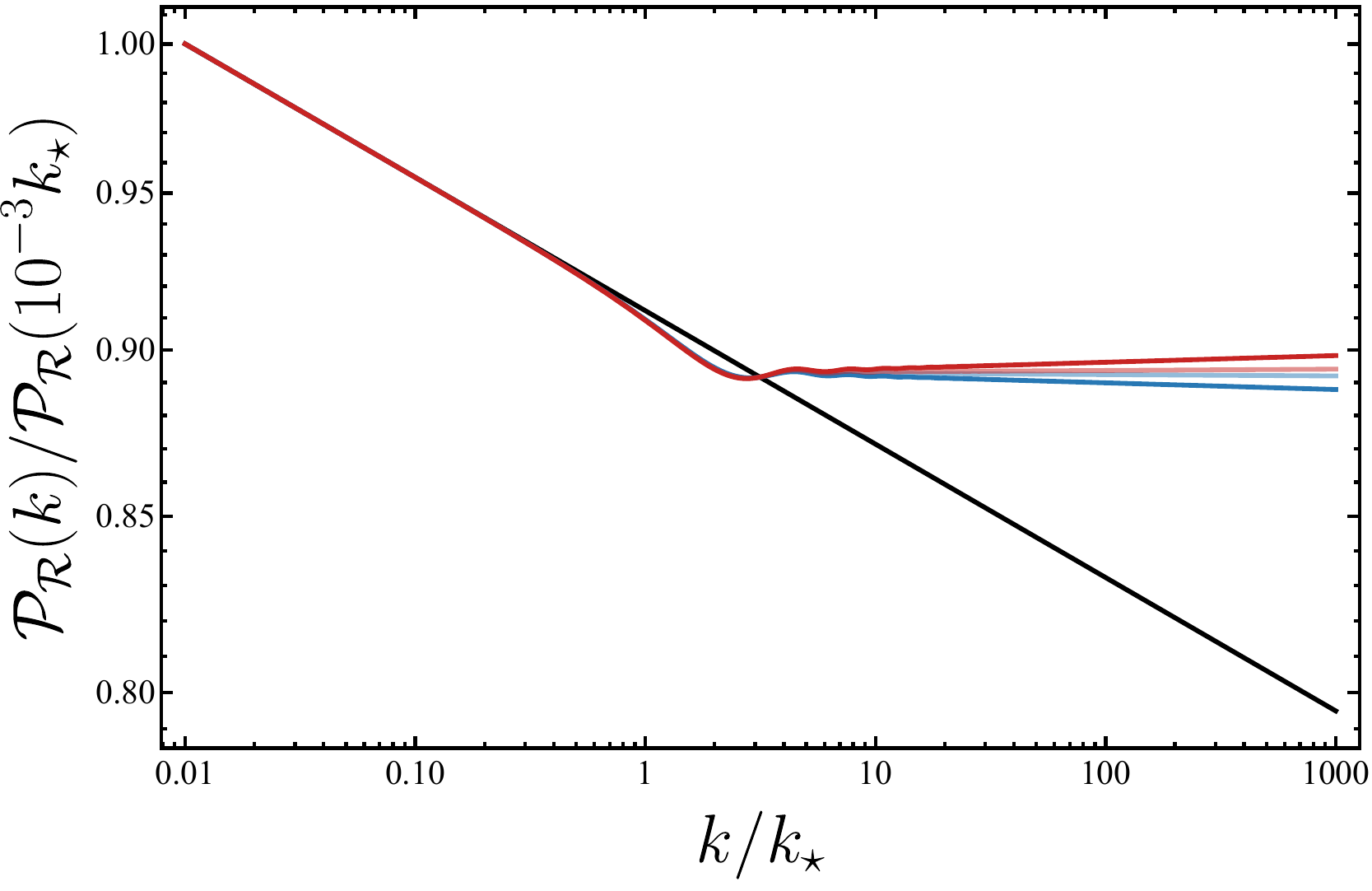}
\begin{center}
\begin{tabular}{ |c||r|r|r| } 
\hline
  $\eta_{II}$ & Top left~~&Top right~~&Bottom~~~\\ 
\hline
 Black & $-10^{-2}$ & $-10^{-2}$ &  $-10^{-2}$ \\ 
 \hline
 Dark blue & $-5\times10^{-1}$ & $-5\times10^{-3}$ &  $-5\times10^{-4}$ \\ 
\hline
 Light blue & $-10^{-1}$  & $-10^{-3}$ & $-10^{-4}$  \\ 
  \hline
 Light red & $10^{-1}$ & $10^{-3}$ & $10^{-4}$\\
 \hline
  Dark red  &$5\times10^{-1}$ & $5\times10^{-3}$ & $5\times10^{-4}$ \\ 
 \hline
\end{tabular}
\end{center}
\caption{Spectra for $R_{\epsilon}^2=1$, $\eta_I=-10^{-2}$ and for various values of $\eta_{II}$ shown in the table.}
\label{powerspectrum}
\end{figure}

\begin{figure}[htbp]
\centering
\includegraphics[width=0.45\textwidth]{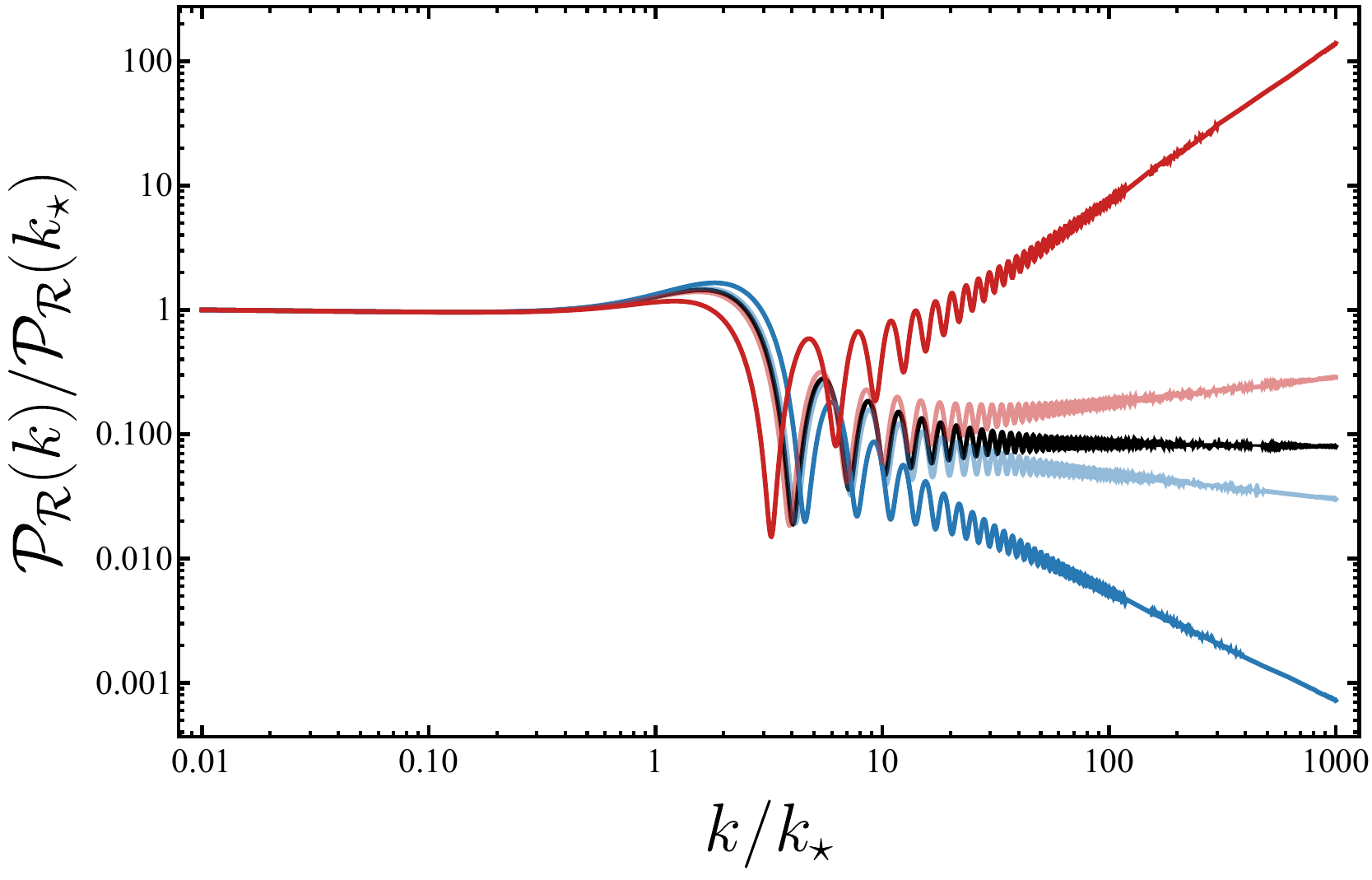}
\qquad
\includegraphics[width=0.45\textwidth]{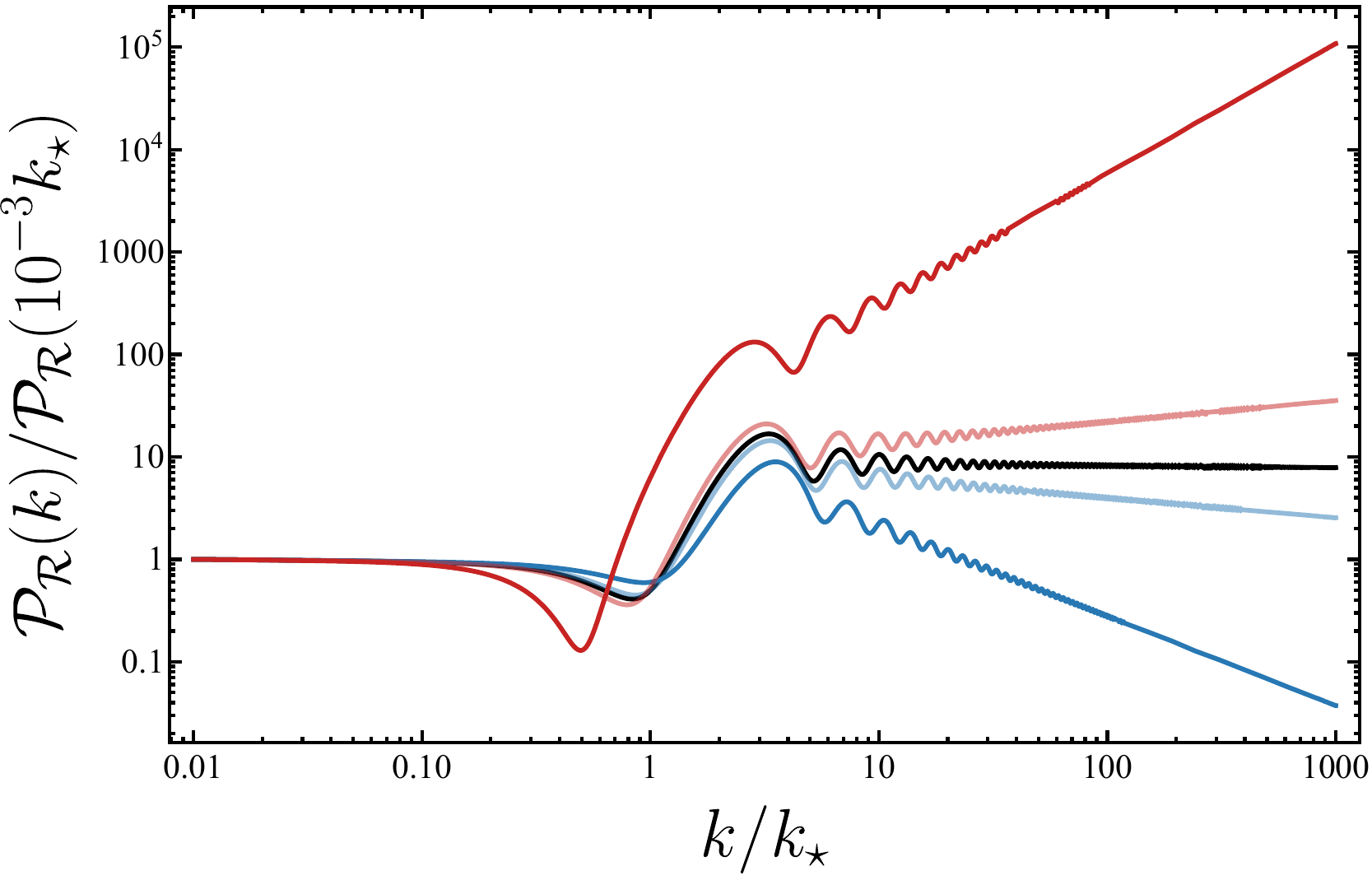}
\caption{Spectra for $\eta_I=-10^{-2}$ and for various values of $\eta_{II}$: $\eta_{II}=-10^{-2}$ (black), $-5\times10^{-1}$ (dark blue), $-10^{-1}$ (light blue), $10^{-1}$ (light red), and $5\times10^{-1}$ (dark red).
The left panel is for $R_\epsilon^2=10$ and the right panel for $R_\epsilon^2=0.1$.}
\label{powerspectrumetachange}
\end{figure}

The analytical result for $\left.\mathcal{P}_{\mathcal{R}}(k)\right|_{\tau\rightarrow0}$ under the constant $H$ approximation is obtained by inserting \eqref{adiabaticvac}  and \eqref{cdana} into \eqref{analyticalpowerspectrum}.
We find it matches the result obtained by numerically solving the perturbation equations very well, as shown in fig.~\ref{anavsnumresult}.
Using the analytical formula for the power spectrum, we plot several characteristic examples in figs.~\ref{powerspectrum} and \ref{powerspectrumetachange}. 
Both figures show the cases with different $\eta$ for fixed $R_\epsilon$.

In the figures above, the shape of the power spectrum varies with the parameter sets we used. In order to understand the features analytically, we divide our detailed discussion into the two parts, (1) Long wavelength modes that exit the horizon before $t_{\star}$, i.e. $k\ll k_\star$, (2) Short wavelength modes that exit the horizon after $t_{\star}$, i.e. $k\gg k_\star$.

\subsection{The long wavelength limit}
Assuming that there was no stage with features prior to stage 1, we set $D_I=0$.
Then expanding the analytical expression for the power spectrum \eqref{analyticalpowerspectrum}  to second order in $X=k/k_\star\ll1$, we find
\begin{align}
  \mathcal{P}_\mathcal{R}(k;\tau_{f})&=\mathcal{A}_\mathcal{R}^{L}
    X^{3-2\nu_{I}}
    {\left|1+\xi_1 X^{2\nu_{I}}+\xi_{2} X^{2}+\mathcal{O}(X^{2+2\nu_{I}})\right|^2},
    \label{IR}
\end{align}
where the amplitude $\mathcal{A}_\mathcal{R}^{L}$ is given by
\begin{align}
    \mathcal{A}_\mathcal{R}^{L}=\left(\frac{2^{-5/2+\nu_I}}{3}\frac{H_\star}{2\pi}\frac{\Gamma(\nu_I)}{\sqrt{\pi}/2}\frac{3+2\nu_{I}}{\sqrt{2\epsilon_{I}}}\right)^2,
\end{align}
and the coefficients $\xi_1$ and $\xi_2$ are given by 
\begin{align}
\label{prcoe}
    &\xi_1\simeq-4^{-\nu_{I}}\frac{\pi(\cot(\nu_{I}\pi)-i)}{\Gamma(1+\nu_{I})\Gamma(\nu_I)}\left(1-\frac{4\nu_{I}}{2\nu_{II}-3+(2\nu_{I}+3)R_{\epsilon}}\right),\\
    &\xi_2\simeq\frac{1}{4(\nu_{II}+1)(\nu_I-1)}\left(2+{\nu_{II}-\nu_{I}}-\frac{4(\nu_{I}+\nu_{II})}{2\nu_{II}-3+(2\nu_{I}+3)R_{\epsilon}}\right).
\end{align}

We note that the curvature perturbation approaches a constant in the superhorizon limit in general, but there are cases when the curvature perturbation evolves in time for a while on superhorizon scales if there is a feature in the potential.
Therefore, the value of the curvature perturbation at (shortly after) the horizon crossing may not be the same as that at the end of inflation. This is the reason why we evaluate the power spectrum at the end of inflation by taking the limit $-k\tau\to0$.

In the very long wavelength limit, $X=k/k_\star\to0$, apparently the leading order term $X^{3/2-\nu_I}$ ($3/2-\nu_I\approx\eta_I$ for $\eta_I\ll1$) dominates the power spectrum, obtaining a slightly tilted power spectrum due to the non-vanishing $\eta_{I}$,
\begin{align}
    \mathcal{P}_{\mathcal{R}}^{{IR}}(k)= \left(\frac{H_\star}{2\pi}\right)^2\left(\frac{k/k_{\star}}{2}\right)^{3-2\nu_{I}}\left(\frac{\Gamma(\nu_{I})}{\sqrt{\pi}/2}\frac{2\nu_{I}+3}{6\sqrt{2\epsilon_{I}}}\right)^2.
\end{align}
Nevertheless, the next to leading order (NLO) terms may dominate the spectrum of near IR modes $k\lesssim k_\star $, unless both $|\xi_1|$ and $|\xi_2|$ happen to be much smaller than unity. 
Let us call the NLO terms, $\xi_1 X^{2\nu_{I}}$ and $\xi_{2}X^{2}$, respectively, T1 and T2.
When either T1 or T2 dominates, the scaling behavior of the near IR modes in the final power spectrum will be strongly tilted.

In order to look into this case in detail, let us first clarify the magnitude of T1 and T2.
When $\nu_{I}<1$, T1 is the term with a lower power-law index, and T2 for $\nu_{I}>1$. The ratio of $\xi_1$ and $\xi_2$ reads
\begin{align}
    \frac{\xi_2}{\xi_1}=\frac{4^{\nu_I-1}\Gamma(\nu_I)\Gamma(1+\nu_I)}{\pi(1+\nu_{II})(\cot(\nu_{I}\pi)-i)}\left(1+\frac{1+\nu_{II}}{1-\nu_I}+\frac{4(\nu_I+\nu_{II})}{4\nu_{I}+(2\nu_{II}-3)(\alpha^{-1}-1)}\right).
    \label{xiratio}
\end{align}
If both $\eta_I$ and $\eta_{II}$ are small ($\ll1$), we obtain $ {\xi_2}/{\xi_1}\approx-1.2i-3.77\eta_{I}-0.08i\eta_{II}$. We see that the effect of $\eta_I$ on the ratio is much stronger than that of $\eta_{II}$.
By inspecting the ratio numerically, we find the $\eta_{II}$ dependence is small even if it is of order unity. 
Hence, when considering the parameter dependence of the ratio $\xi_2/\xi_2$, we may safely neglect the $\eta_{II}$ contribution and focus on the $\eta_I$ dependence. 

\begin{figure}[htbp]
\centering
\includegraphics[width=0.8\textwidth]{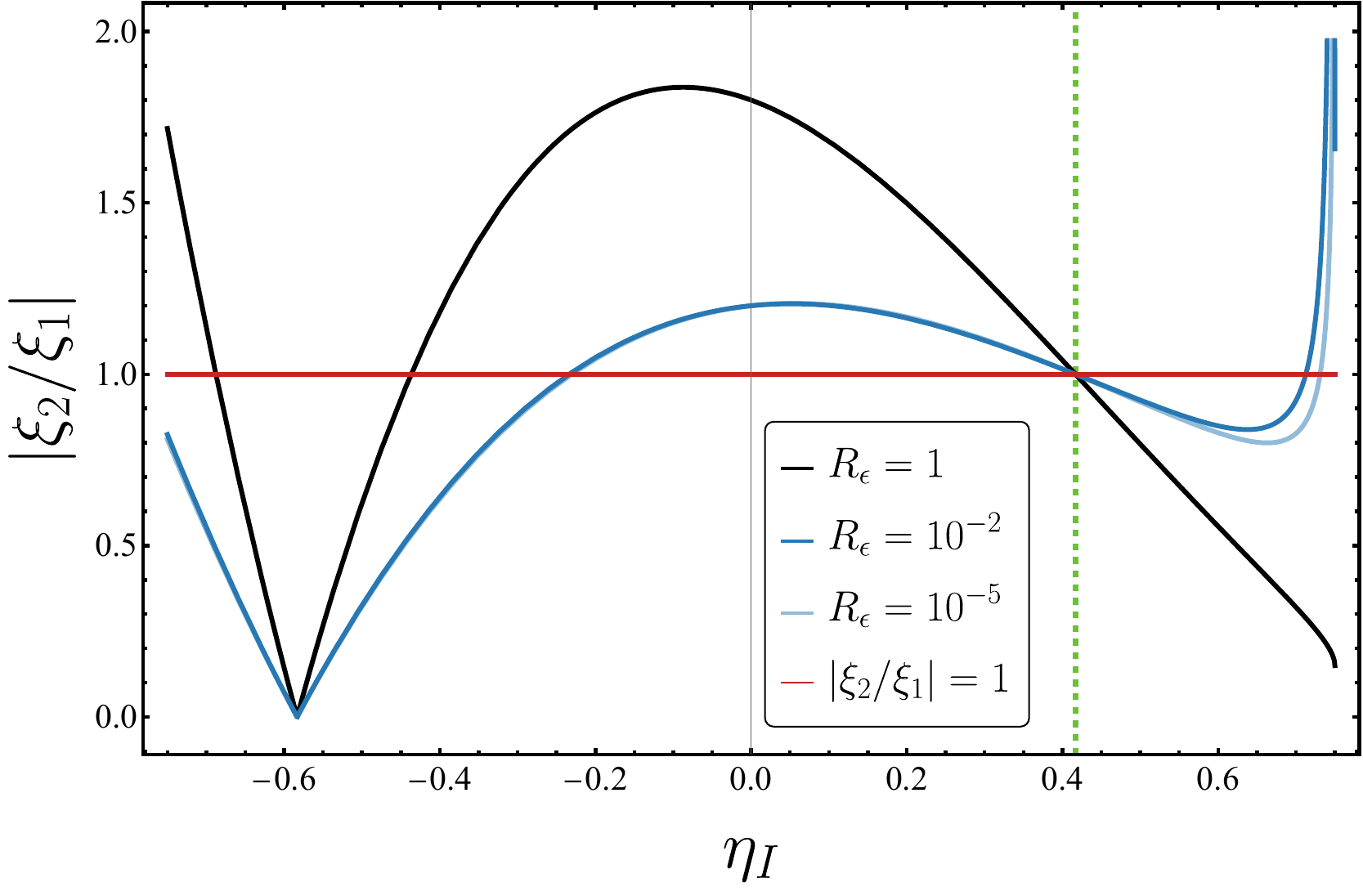}
\caption{\textbf{The dependence of $\xi_2/\xi_1 $ to $\eta_{I}$.} For the black, blue and light blue solid lines, we take $R_\epsilon=1,10^{-2},10^{-5}$ to plot the expression given by \eqref{xiratio}. The green vertical line indicates $\eta_{I}=5/12$, and the horizontal red line shows the $|\xi_2/\xi_1|=1$, they divide the figure into four parts, for parameters lie in the top-left part, the $\xi_2$ term serves as next to leading order term with the biggest coefficient among the higher order terms, thus dominates the power spectrum for $|\alpha-1|<1$ cases.  Whilst for parameters lie in the bottom right corner, the $\xi_1$ term dominates. }
\label{x2ox1}
\end{figure}

We plot it in fig.~\ref{x2ox1}, where the limit $|\eta_{II}|\to0$ is taken.
As one can notice, there are two distinct values of $\eta_I$, i.e., $|\xi_2/\xi_1|=1$ at $\eta_I=5/12$ and  $|\xi_2/\xi_1|=0$ at $\eta_I=-7/12$, independent of the value of $R_\epsilon$.
For $\eta_I<-7/12$, the T2 contribution increases as $\eta_I$ decreases. For $-7/12<\eta_I<5/12$, the T2 contribution is maximized around $\eta_I=0$. For $\eta_I>5/12$, the T2 contribution may become dominant depending on the value of $R_\epsilon$.
This $R_\epsilon$ dependence can be seen by taking the limit $\eta_{I}\to 3/4$. We obtain $\xi_2/\xi_1\rightarrow7/20-1/(5R_{\epsilon})$. This is smaller than unity for large $R_\epsilon$, but exceeds unity for $R_\epsilon<4/27$. 

\begin{figure}[htbp]
\centering
\includegraphics[width=0.45\textwidth]{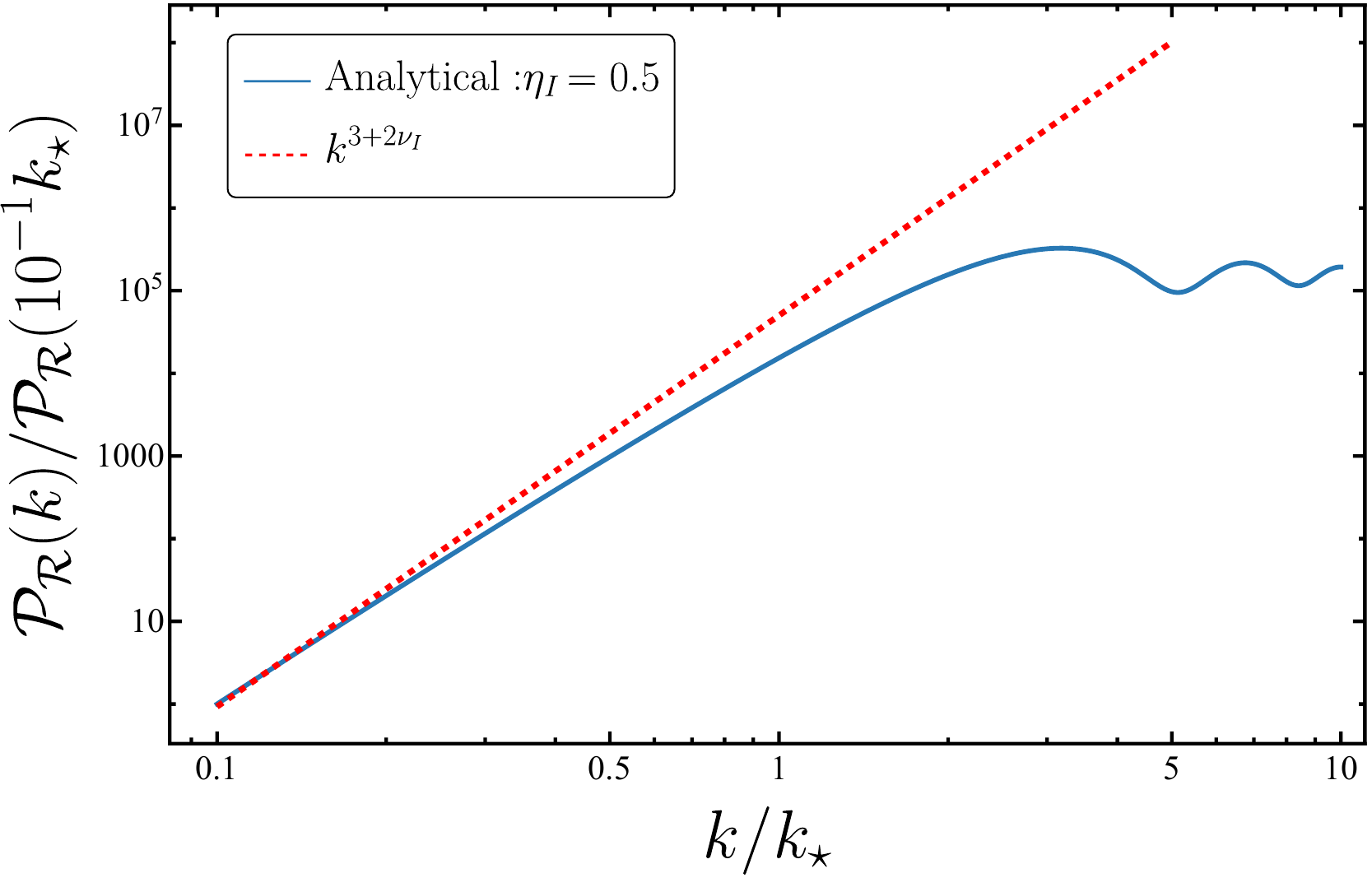}
\includegraphics[width=0.45\textwidth]{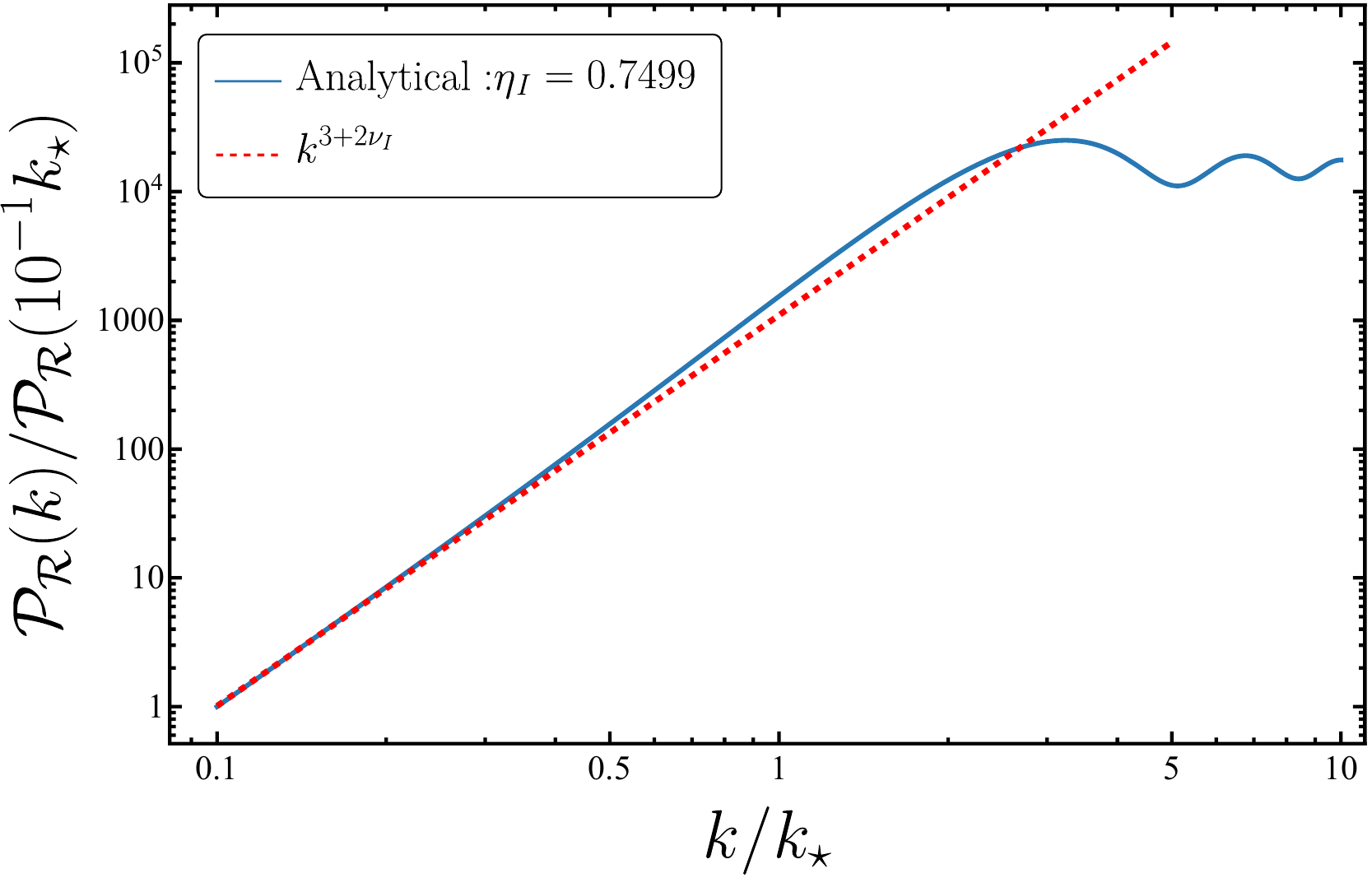}
\caption{\textbf{The normal power spectrum V.S. sharpened power spectrum. }We take $R_\epsilon=10^{-5}$ for both graphs. }
\label{sharpened}
\end{figure}

In the above, we ignored the possible higher-order contribution. To estimate them analytically is beyond the scope of the current paper, but as far as the range of parameters we consider is concerned, a good fit with the numerical results suggests that the higher orders are unimportant in most cases.
Nevertheless, when $|\xi_2|\gg1$, the higher order contribution may not be negligible near $k\sim k_\star$. In fact, the numerical result we obtained shows a slightly sharpened growth feature near $k_\star$ as depicted in fig.~\ref{sharpened}. 


Let us turn back to the case T2 dominates, and focus on the magnitude of $\xi_2$. 
As we have discussed, T2 dominance happens for a relatively small $\eta_{I}\sim \mathcal{O}(0.1)$.
By expanding $\xi_2$ in the limit $\eta_{I}$ and $\eta_{II}$ are small, we obtain
\begin{align}
    \xi_2=-\frac{\eta_{II}(\eta_{II}-2)+\alpha(6+\eta_{I}(\eta_{II}-2)-\eta_{II}^2)}{(1-\alpha)(2\eta_{I}-1)\eta_{II}(2\eta_{II}-5)}\approx\frac{2(\eta_{II}+\alpha(\eta_{I}-3))}{5\eta_{II}(1-\alpha)}.
\end{align}
Apparently, the term $1-\alpha$ in the denominator determines the behavior of the power spectrum for near-IR modes ( $k\lesssim k_\star$ ). 
To be specific, T2 starts to dominate the full power spectrum for modes in the range,
\begin{align}
1> \frac{k}{k_\star}>\frac{1}{{\sqrt{|\xi_2|}}}\approx\sqrt{\left|\frac{5\eta_{II}(1-\alpha)}{2(\eta_{II}+\alpha(\eta_{I}-3))}\right|},
    \label{consonk}
\end{align}
when $|\xi_2|\gg1$, which is realized either for $|\delta\alpha|\ll 1$ (where $\delta\alpha\equiv\alpha-1$). 
In this case, the right inequality in \eqref{consonk} can easily be fulfilled even by very small k, since it simplifies to $k/k_\star\gtrsim\sqrt{5\eta_{II}|\delta\alpha|/6}$,
where we used the fact that $\alpha\propto (3-2\nu_{II})>0$ implies $\eta_{II}>0$ according to \eqref{alpha}.

In the limit both $\eta_I$ and $\eta_{II}$ are very small, while $R_\epsilon$ is not so small, we have $\xi_2=2(R_\epsilon-1)/(5R_\epsilon)$, which becomes smaller than $-1$ when $R_\epsilon<2/7$. Note that \eqref{approxalpha} implies $|\alpha|\ll1$ in this case.
In this limit, T2 dominates since T1 is proportional to $X^{2\nu_I}\sim X^3$ while T2 to $X^2$. Thus, a dip may appear at $X=k/k_\star<1$ when $R_\epsilon<2/7$, and the spectrum is dominated by T2, hence ${\cal P}_{\cal R}(k)\propto k^4$ at $k/k_\star<1$. 

Another interesting finding is that the sign of $\xi_{2}$, which determines the existence or nonexistence of a dip in the power spectrum, depends on whether $\alpha$ is greater or smaller than unity.
For the case $|\delta\alpha|=|\alpha-1|\ll1$, we obtain a simplified form of $\xi_2\approx{6}/({5\eta_{II}\delta\alpha})\gg1$. Thus, the sign of $\xi_{2}$ directly depends on the sign of $\delta\alpha$. 

\begin{figure}[htbp]
\centering
\includegraphics[width=0.8\textwidth]{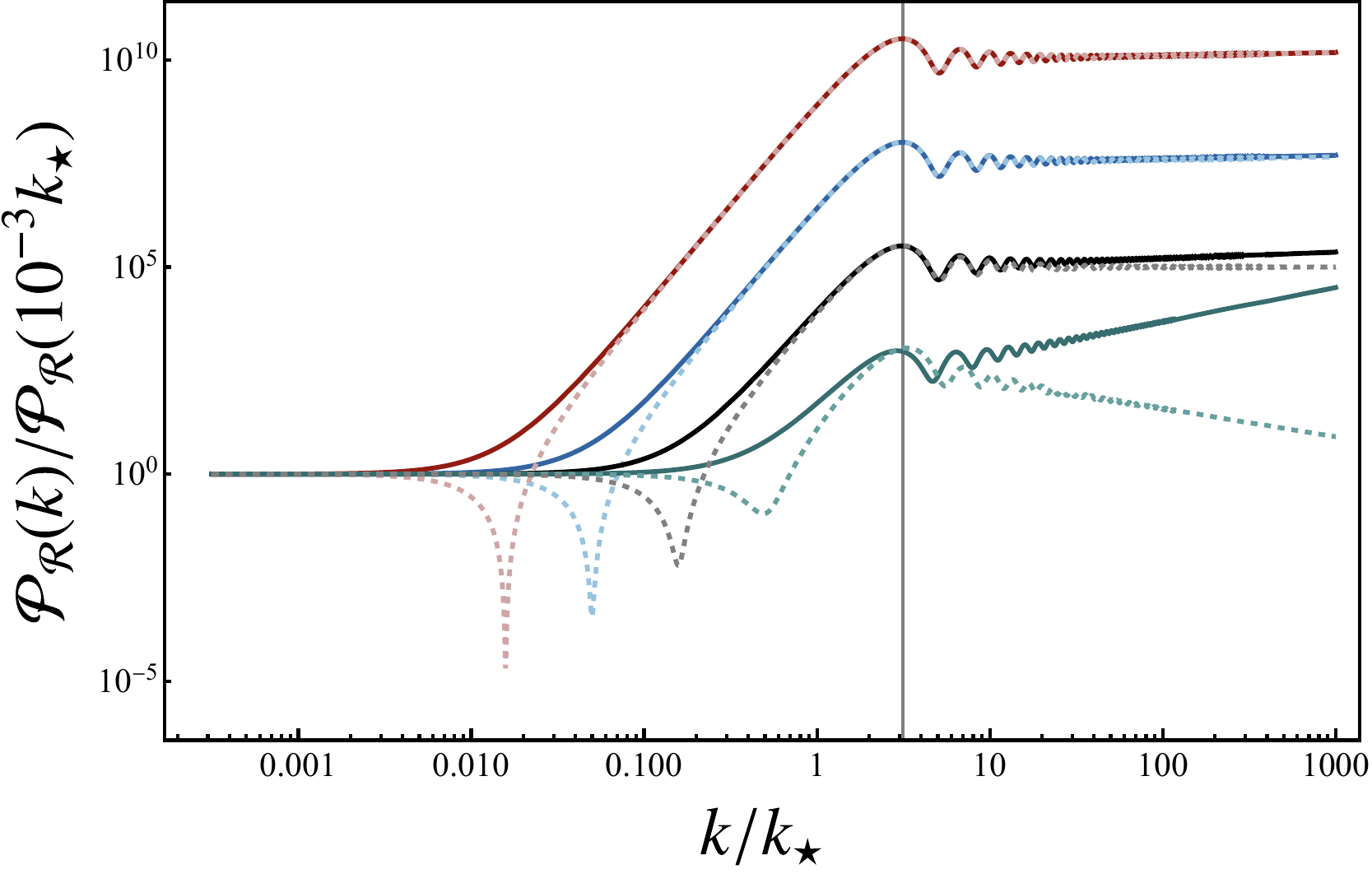}
\caption{\textbf{The enhancement of power spectrum for $\delta\alpha$ with different signs. } The figure is given by  \eqref{analyticalpowerspectrum},  by setting $R_{\epsilon}=10^{-2}$, $\eta_{I}=-1\times10^{-10}$ and $D_{I}=0$. For the red, blue, black, and green solid (dashed) lines, we take $\delta\alpha=+(-)10^{-2},+(-)10^{-1},+(-)1,+(-)10$ to show the alpha dependence of the shape of the power spectrum.
The gray vertical line indicates $k/k_\star=\pi$, which is the analytical approximation of the peak position discussed in sec.~\ref{shortwave}.}
\label{uturn}
\end{figure}

We have $\xi_2<0$ if $\delta\alpha<0$ ($\alpha<1$). 
In this case, a dip exists, and the position of the scale at which the dip appears is given by 
\begin{align}
    \frac{k_{\mathrm{dip}}}{k_\star}=|\xi_2|^{1/2}=\sqrt{\left|\frac{(1-\alpha)(2\eta_{I}-1)\eta_{II}(2\eta_{II}-5)}{\eta_{II}(\Delta\eta_{II}-2)+\alpha(6+\eta_{I}(\eta_{II}-2)-\eta_{II}^2)}\right|}\approx\sqrt{\left|\frac{5\eta_{II}(1-\alpha)}{2(\eta_{II}+\alpha(\eta_{I}-3))}\right|}.
\end{align}
On the other hand, if $\delta\alpha>0$ ($\alpha>1$), we have $\xi_2>0$, and the dip ceases to exist (as we show in fig.~\ref{uturn}).
It is worth noting here that since $\alpha>1$ is the condition for the disappearance of a dip, it means that the attractor phase is realized only after the inflaton passes through the minimum, climbs up the potential, stops moving and makes a u-turn toward the minimum, as discussed in sec.~\ref{sec:background}. 

It is also worth mentioning that, in our model, the steepest growth of the power spectrum can exceed the "steepest growth" of $k^4$ if we tune the parameters. According to the \eqref{IR}, assuming both $\xi_1$ and $\xi_2$ are of the same order, the spectra index $n_\star$ near $k_\star$ is given by 
\begin{equation}
    n_\star=\min[3+2\nu_I,7-2\nu_I]\,. 
\label{spectralindex}
\end{equation}
This takes the maximum value of 5 at $\nu_I=1$ ($\eta_I=5/12$) where both indices degenerate. When the index $\nu$ of the Hankel function takes an integer value, the solution near the origin starts to contain a logarithmic function. This gives rise to a logarithmic dependence of the spectral index in $k$. Thus the
resulting spectral shape is becomes $k^5(\ln k)^2$. 

One may then wonder if there could be a case when $\xi_2=0$ and $|\xi_1|>1$ for $\nu_I>1$. If this happens the spectral index would be given by $3+2\nu_I$ and it would exceed 5. In order to examine this possibility, we performed a perturbative analysis by setting $\xi_2=0$ and $\nu_I=1+\delta$ ($\delta>0$). We found that
$\xi_1$ can never exceed unity. Thus we conclude that the steepest growth rate in our model is $k^5(\ln k)^2$ \cite{Carrilho:2019oqg}.


\begin{figure}[htbp]
\centering
\includegraphics[width=0.8\textwidth]{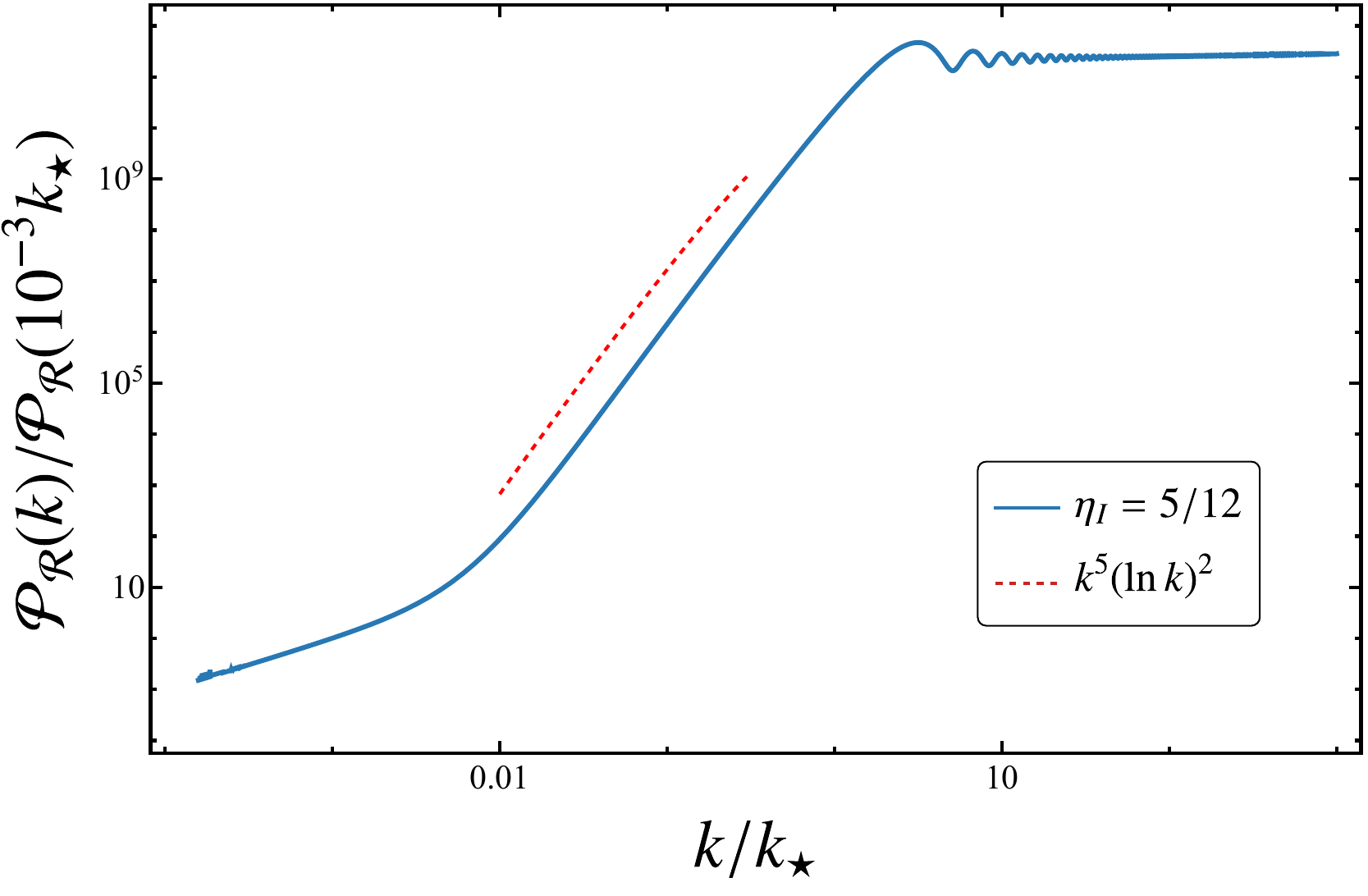}
\caption{\textbf{The steepest growth of $k^5(\ln{k})^2$}. In this plot, we take $\eta_{I}=5/12$, $R_{\epsilon}=10^{-5}$.}
\label{steep}
\end{figure}

As the appearance of a logarithmic function is a property of the Hankel function with an integer index,
one might suspect that the $(\ln k)^2$ feature in the power spectrum could appear not only when $\nu_I=1$ but when $\nu_I=2,\,3,\cdots$. However, Eq.~\eqref{spectralindex} tells us that the T2 term always donimates, hence renders the logarithmic term subdominant.


It may be noted that our discussion of the steepest growth $k^5(\ln{k})^2$ is restricted to the case of the adiabatic vacuum. As recently found in \cite{Cielo:2024poz}, for the $\alpha$ vacuum, a growth of $k^6$ is possible for some certain choice of the Bogoliubov coefficient. In our case, the $\alpha$ vacuum corresponds to the choice of non-vanishing $D_I$ in Eq.~\eqref{adiabaticvac}. 
We leave the analysis for the $\alpha$-vacuum in the generalized piecewise potential in the future work. 

\subsection{The short wavelength limit}
\label{shortwave}
In the limit $X=k/k_\star\gg 1$, the final amplitude of the curvature perturbation can be expanded to second order as
\begin{align}
    \mathcal{P}_\mathcal{R}(k;\tau_{f})=\mathcal{A}_\mathcal{R}^{S}\left|{X^{\frac{3}{2}-\nu_{II}}}+\Omega_1(X){X^{\frac{1}{2}-\nu_{II}}}+\Omega_2(X){X^{-\frac{1}{2}-\nu_{II}}}+\mathcal{O}(X^{-\frac{3}{2}-\nu_{II}})\right|^2,
   \label{UV}
\end{align}
where the amplitude $\mathcal{A}_\mathcal{R}^{S}$ is given by
\begin{align}
    \mathcal{A}_\mathcal{R}^{S}&=\left(\frac{2^{-5/2+\nu_{II}}}{3}\frac{H_\star}{2\pi}\frac{\Gamma(\nu_{II})}{\sqrt{\pi}/2}\frac{4\nu_{II}}{(1-\alpha)\sqrt{2\epsilon_{II}}}\right)^2
    \nonumber\\
    &=\mathcal{A}_\mathcal{R}^{L}\,\left(\frac{2^{\nu_{II}}\Gamma(\nu_{II})}{2^{\nu_I}\Gamma(\nu_{I})}\frac{4\nu_{II}}{(-3+2\nu_{II})+R_\epsilon(3+2\nu_{I})}\right)^2\,.
    \end{align}
Thus the enhancement of the power spectrum in the short wavelength regime is estimated as
\begin{align}
    \frac{\mathcal{P}_{\mathcal R}(k_{UV})}{\mathcal{P}_{\mathcal R}(k_{IR})}=\frac{  \mathcal{A}_\mathcal{R}^{S}}{  \mathcal{A}_\mathcal{R}^{L}}&=\left(\frac{2^{\nu_{II}}\Gamma(\nu_{II})}{2^{\nu_{I}}\Gamma(\nu_{I})}\right)^2\left(\frac{4\nu_{II}}{(-3+2\nu_{II})+R_\epsilon(3+2\nu_{I})}\right)^2
    \nonumber\\&=
  \left(\frac{2^{\nu_{II}}\Gamma(\nu_{II})}{2^{\nu_{I}}\Gamma(\nu_{I})}\right)^2\frac{1}{(1-\alpha)^2}\left(\frac{4\nu_{II}}{(3+2\nu_{I})R_{\epsilon}}\right)^2.
    \label{amplification}
\end{align}
This slightly differs from the estimate \eqref{prestim} in the small $\eta$ limit (ie, both $\eta_I$ and $\eta_{II}$ are small), in which the factor involving the Gamma functions is absent. Here the expression is valid for any models with non-negligible $\eta$. 
The functions $\Omega_1$ and $\Omega_2$ in \eqref{UV} are given by 
\begin{align}
  \Omega_1  =&\frac{1}{4}\left[\left(e^{2i(X-\nu_{II}\pi)}+i\right)(3+2\nu_{I})(R_{\epsilon}-1)-2i(\nu_{II}^2-\nu_{I}^2)\right],\\
    \Omega_2=&\frac{1}{16}ie^{2i(X-\nu_{II}\pi)}\left[(3+2\nu_1)(2\nu_{I}^2+2\nu_{II}^2-1)(R_{\epsilon}-1)-4(\nu_{II}^2-\nu_{I}^2)\right]
    \nonumber\\
    &+\frac{1}{8}(\nu_{II}^2-\nu_{I}^2)\left[(3+2\nu_1)(R_{\epsilon}-1)-(\nu_{II}^2-\nu_{I}^2)\right].
\end{align}
As they contain the oscillatory function $e^{i2X}$, and an oscillating feature appears in the power spectrum when $R_\epsilon\neq 1$, as seen in Figs.~\ref{anavsnumresult}, \ref{powerspectrum} and \ref{powerspectrumetachange}. 

In the regime of $R_\epsilon\ll 1$, $\Omega_1$ and $\Omega_2$ are approximately of the same order. 
Since $\nu_I$ and $\nu_{II}$ are both positive, the $\Omega_1x^{1/2-\nu_{II}}$ term dominates the oscillation, 
with $\Omega_1$ approximately given by
\begin{align}
      \Omega_1  \approx&-\frac{(3+2\nu_{I})}{4}\Bigl(1+\exp\bigl[i\left(2X-2\nu_{II}\right)\bigr]\Bigr)\,.
\end{align}
The first peak gives the local maximum for modes $k>k_\star$. 
In the small $\eta$ limit, by solving $d|\Omega_1|/dx=0$, we find it is located at \cite{Pi:2022zxs}
\begin{align}
    X_{\mathrm{peak}}\approx\frac{\pi}{4}(1+\nu_{I}+\nu_{II})\sim\pi\,,
\end{align}
as shown in Fig.~\ref{uturn}. 
In the small wavelength limit, the leading order term in \eqref{UV} implies a slightly tilted spectrum proportional to $k^{3-2\nu_{II}}$. Some examples of the case with significant tilts are given in Figs.~\ref{powerspectrum} and \ref{powerspectrumetachange}. 

To conclude this section, let us summarize the $\eta$ dependence of the shape of the power spectrum. It appears in four ways:
\\
\begin{itemize}
    \item  The slope of the far IR (UV) part of the power spectrum is determined by $\eta_I$ $(\eta_{II})$ provided that it is much larger than $\epsilon$ (ie, $\epsilon_X\ll|\eta_X|$ for $X=I,\,II$). It is proportional to $k^{3-2\nu_{X}}$.
    \item The amplitude of the enhancement is determined by parameter $\alpha$ defined by Eq.~\eqref{alpha}, which is a function of $\epsilon_{X}$ and $\eta_{X}$ ($X=I,\,II$), that controls the state of inflaton at the end of the off-attractor stage.
As given by \eqref{amplification}, the small scale power spectrum is strongly enhanced when $|\alpha-1|\ll 1$.
\item The scaling behavior of the enhanced power spectrum for modes $k\lesssim k_\star$ is modified by $\eta_{I}$, resulting in the steepest growth of $k^5(\ln k)^2$ for $\nu_I=1$ or equivalently $\eta_{I}=5/12$. 
\item The dip at $k<k_*$ and the peak at $k>k_\star$ of the power spectrum depend also on the values of $\eta_X$. When $\alpha>1$, which may happen if $\eta_{II}>0$, the dip disappears. The peak position is well approximated by $k\approx \pi k_\star$ when both $\eta_X$ ($X=I,\,II$) are small.
\end{itemize}

\section{The non-Gaussianity: $\delta N$ calculation}
\label{sec: The non-Gaussianity: deltaN calculation}
In sec.~\ref{sec:background}, we have discussed the background evolution of  inflaton with the potential with the form presented as ~\eqref{potential}.
In this section, applying the $\delta N$ formalism to the background solutions given there, we discuss the non-Gaussianity in this model.

The $\delta N$ formalism is based on the separate universe approach that a comoving region greater than the Hubble horizon scale evolves like a homogeneous and isotropic universe at leading order in the spatial gradient expansion. It offers an intuitive and efficient method for linking the field fluctuations $\delta\phi$ with the curvature perturbation at the end of inflation by $\mathcal{R} = \delta N$. 
Here, $N$ represents the e-folding number of a fiducial patch of the universe, calculated backward in time from the end of inflation as $N \equiv n_{\rm end} - n$, where $\dd n \equiv H\dd t$, to an epoch when the scalar field $\phi$ and its velocity $\pi=-d\phi/dn$ takes given values, i.e., $N=N(\phi,\pi)$. 
The $\delta N$ is then given by the difference in the e-folding number between the two different initial conditions $\delta N=N(\phi+\delta\phi, \pi+\delta\pi)-N(\phi,\pi)$, where $\delta\phi$ and $\delta\pi$ are given by the field fluctuations on the flat slicing at or after the horizon crossing.

Usually $\delta\phi$ approaches the adiabatic limit, $\delta\phi\propto \pi$ and $\delta\pi\to0$. In this case, one can use the values of $\delta\phi$ at the horizon crossing and setting $\delta\pi=0$ as the initial condition. This naturally introduces the scale-dependence of $\delta N$ determined by the horizon crossing time of the comoving scale. 

However, it is now well understood that there can be non-trivial evolution of the field fluctuations even after horizon crossing if there exists a feature in the potential that induces transitional behavior of the scalar field \cite{Leach:2001zf, Jackson:2023obv}, like the case we consider in the current paper. 
In such a case, it is known that the next leading order correction to $\delta\phi$ in gradient expansion must be taken into account in order to evaluate the curvature perturbation spectrum properly if we use the horizon crossing value of $\delta\phi$.
An extension of the $\delta N$ formalism that includes this correction has been done recently in  \cite{Artigas:2024xhc}. 

In this paper, in order to avoid these issues, we apply the $\delta N$ formula only at epoch sufficiently far outside the horizon when the scalar field perturbation has relaxed to the attractor limit at some late time $n=n_{\rm attr}>n_\star$,
under the assumption that the scalar field perturbation can still be treated in linear theory until $n=n_{\rm attr}$.
Then provided that the scalar field perturbation follows the classical equations of motion, using the fact that $\delta N$ is independent of the time at which it is evaluated, one may pretend that one can evaluate $\delta N$ at horizon-crossing $n=n_k$, not with the actual values of  $\delta\phi$ and $\delta\pi$ at horizon-crossing, but with the values determined by those evolved backward in time from $n_{\rm attr}$ to $n_k$. With this prescription, we make sure that the non-Gaussian properties evaluated by using the $\delta N$ formula are correct even if we take the initial slice to be the one at horizon-crossing.

In passing, we note that there is a subtlety in the evolution of $\delta\phi$ on superhorizon scales that has not been explicitly stated in most literature. 
Usually, the evolution of $\delta\phi$ is assumed to be linear.
As in the case of the first or second stage of our model, this is true as long as the equation of motion is a linear second-order differential equation. 
However, if there is a non-trivial feature like the sudden change in the slope and/or the effective mass, its effect is the same as the case of a highly nonlinear potential for which the nonlinear corrections to the evolution of $\delta\phi$ is non-negligible. Thus if we evolve $\delta\phi$ from the first stage to the second stage on superhorizon scales, $\delta\phi$ at the first stage is perfectly Gaussian, but that at the second stage contains the nonlinear corrections due to the transition, hence is non-Gaussian in general. A detailed discussion on this point will be discussed in the subsection at the end of this section.

\subsection{Analytical formulas for $\delta N$}
First, let us summarize the qualitative behavior of $\delta N$ as one varies the comoving scale.   
Broadly speaking, they can be categorized into three cases depending on the horizon crossing time. Perturbations corresponding to larger scales (long-wavelength modes) exit the Hubble horizon earlier:
\begin{itemize}
    \item Case A: Super-long-wavelength modes $n_\star-n_k\gg1$. An approximate analytical expression for $\delta N$  is given later by Eq.~\eqref{casea}.
    \item Case B: Modes that exit the horizon shortly before the transition point $n_k\lesssim n_\star$. An approximate analytical expression is given by Eq.~\eqref{near phi_* modes 1}.
    \item Case C: Modes that exit the horizon after the transition point $n_k>n_\star$.
    An approximate analytical expression is given by Eq.~\eqref{near phi_* modes 2}.
\end{itemize}

Apparently, Case A and Case B differ from Case C in that the modes exit the horizon before the transition $n=n_\star$ for the former, while they exit the horizon after $n=n_\star$ for the latter. 
Thus, the modes undergo a two-stage evolution for Case A and Case B ($n_k<n_\star$), hence $N = N_I+N_{II}$, where $N_I=n_{\star} - n_k$ and $N_{II}=n_{\rm end}-n_\star$.
In contrast, for Case C ($n_k>n_\star$), there is only a single stage, $N = n_{\text{end}} - n_k$. 
The analytical results corresponding to these three cases can be obtained by applying the technique developed in \cite{Pi:2022ysn}. 
A detailed numerical comparison is provided in sec.~\ref{appendix: numerical delta N}.
Below we recapitulate the formulas in \cite{Pi:2022ysn}.

By manipulating the background solution \eqref{eq:phi solution}, we have
\begin{align}\label{trajectoies with N}
    \frac{\pi_{X} + \lambda_{X,+} \widetilde{\phi}_{X}}{\pi_{\star} + \lambda_{X,+} \widetilde{\phi}_{X,\star}} &= e^{-\lambda_{X,-}(n-n_{\star})}\,,\\
    \label{trajectoies with Nm}
    \frac{\pi_{X} + \lambda_{X,-} \widetilde{\phi}_{X}}{\pi_{\star} + \lambda_{X,-} \widetilde{\phi}_{X,\star}} &= e^{-\lambda_{X,+}(n-n_{\star})} \,,
\end{align}
where $X=I$ or $II$, and $\widetilde{\phi}_X=\phi_X-\phi_{X,0}$ ($\phi_{X,0}=\phi_\star-\sqrt{2\epsilon_X}/\eta_X$) as before.
Using either one of these expressions, we obtain $n - n_{\star}$ expressed in terms of  $(\widetilde{\phi}, \pi)$ and $\pi_{\star}$ as
\begin{align}\label{e-folds number for single stage}
    n-n_{\star} = -\frac{1}{\lambda_{X,\pm}} \ln\left(\frac{\pi_{X} + \lambda_{X,\mp} \widetilde{\phi}_{X}}{\pi_{\star} + \lambda_{X,\mp} \widetilde{\phi}_{X,\star}}\right) ~.
\end{align}
Furthermore, by eliminating $n-n_{\star}$ from Eqs.~\eqref{trajectoies with N} and \eqref{trajectoies with Nm}, we can derive the relateion that determins $\pi_{\star}$ as a function of $(\widetilde{\phi}_{X}, \pi_{X})$,
\begin{equation}\label{logarithmic duality}
    \begin{aligned}
       {\lambda_{X,-}} \ln \left( \frac{\pi_{X} + \lambda_{X,-} \widetilde{\phi}_{X}}{\pi_{\star} + \lambda_{X,-} \widetilde{\phi}_{X,\star}} \right) &= {\lambda_{X,+}} \ln\left( \frac{\pi_{X} + \lambda_{X,+} \widetilde{\phi}_{X}}{\pi_{\star} + \lambda_{X,+} \widetilde{\phi}_{X,\star}} \right).
    \end{aligned}
\end{equation}
This equality ensures that the result in Eq. \eqref{e-folds number for single stage} is consistent regardless of whether the plus or minus sign is used, leading to the \textit{logarithmic duality} \citep{Pi:2022ysn}. 

Following the procedure in \citep{Pi:2022ysn}, the calculation of the $\delta N$ formalism is straightforward by subtracting Eq. \eqref{e-folds number for single stage} from a perturbed version where $\widetilde{\phi}_{X} \to \widetilde{\phi}_{X} + \delta {\phi}_{X}$, $\pi_{X} \to \pi_{X} + \delta \pi_{X}$, and $\pi_{\star}(\phi,\pi) \to \pi_{\star} + \delta\pi_{\star}$.\footnote{ 
Note that $\phi_{\star}$ is merely a parameter of the model that determines the transition point. Hence it is not perturbed.}
We obtain
\begin{align} \label{delta N-Nstar}
\delta(n - n_{\star})=-\frac{1}{\lambda_{X,\pm}} \ln\left[ 1 + \frac{\delta \pi_{X} + \lambda_{X,\mp} \delta\phi_{X}}{\pi_{X} + \lambda_{X,\mp} \widetilde{\phi}_{X}} \right] + \frac{1}{\lambda_{X,\pm}} \ln\left[ 1 + \frac{\delta \pi_{\star}}{\pi_{\star} + \lambda_{X,\mp} \widetilde{\phi}_{X,\star}} \right]\,,
\end{align}
where $\delta \pi_{\star}(\phi,\pi)$ is determined by the perturbed version of Eq. \eqref{logarithmic duality},
\begin{equation}\label{perturbation of logarithmic duality}
    \begin{aligned}
         &\left( 1 + \frac{\delta\pi_{\star}}{\pi_{\star} +\lambda_{X,-} \widetilde{\phi}_{X,\star}} \right)^{ {\lambda_{X,-}}} \left( 1 + \frac{\delta\pi_{\star}}{\pi_{\star} +\lambda_{X,+} \widetilde{\phi}_{X,\star}} \right)^{ {-\lambda_{X,+}}} \\ 
         &\qquad = \left( 1 + \frac{\delta\pi_{X} + \lambda_{X,+} \delta\phi_{X}}{\pi_{X} + \lambda_{X,+} \widetilde{\phi_{X}}} \right)^{-\lambda_{X,+}} \left( 1 + \frac{\delta\pi_{X} + \lambda_{X,-} \delta\phi_{X}}{\pi_{X} + \lambda_{X,-} \widetilde{\phi_{X}}} \right)^{\lambda_{X,-}}\,,
    \end{aligned}
\end{equation}
which also guarantees the equivalence between the plus and minus signs of Eq. \eqref{delta N-Nstar}.

For $n_k<n_\star$, the evolution at the second stage $n_{\star} < n$ is totally determined by the initial condition at $n=n_\star$, namely, $(\phi_{\star},\pi_{\star})$. 
Thus for $n_k<n_\star$, setting $n=n_k$ and $X=I$ in \eqref{delta N-Nstar} gives $N_I=n_\star - n_k$, and setting $n=n_{\rm end}$ and $X=II$ gives $N_{II}=n_{\rm end}-n_\star$.
On the other hand, for $n_k>n_\star$, $N$ is given by setting $n=n_k$ and $X=II$, and replacing $n_\star$ by $n_{\rm end}$ in \eqref{delta N-Nstar}.
Perturbing thus obtained expressions, we obtain the curvature perturbation at the end of inflation.

For the long-wavelength modes that exit the horizon before the transition (i.e., Case A and B) the comoving curvature at the end of inflation is given by
\begin{align}
    \mathcal{R} \equiv \delta N = \delta N_I +\delta N_{II}\,,
    \label{general R for long wavelength mode}
\end{align}
where 
\begin{align}
    \delta N_I
    &=\delta( n_{\star} - n_k )
 \nonumber\\
 &=\frac{1}{\lambda_{I,\mp}}\left[ \ln\left( 1 + \frac{\delta \pi_{I} +\lambda_{I,\pm} \delta{\phi}_{I}}{\pi_{I} +\lambda_{I,\pm} \widetilde{\phi}_{I}} \right) -\ln\left( 1 + \frac{\delta \pi_{\star}}{\pi_{\star} +\lambda_{I,\pm} \widetilde{\phi}_{I,\star}} \right)\right]\,,
\\
   \delta N_{II}&=\delta ( n_{\text{end}} -  n_{\star} )
   \nonumber\\
   &= \frac{1}{\lambda_{II,\mp}}\left[\ln\left( 1 + \frac{\delta \pi_{\star}}{\pi_{\star} + \lambda_{II,\pm}\widetilde{\phi}_{II,\star}} \right) -\ln\left( 1 + \frac{\delta \pi_{\text{end}}}{\pi_{\text{end}} + \lambda_{II,\pm}\widetilde{\phi}_{\text{end}}} \right)\right].  
\end{align}
In the above, $\delta\pi_\star$ is a function of ($\delta\phi,\delta\pi$)
implicitly determined by Eq.~\eqref{perturbation of logarithmic duality} and $\delta\pi_{\rm end}$ is by 
\begin{equation}\label{delta pi_end}
    \begin{aligned}
                 &\left( 1 + \frac{\delta \pi_{\star}}{\pi_{\star} + \lambda_{II,-}\widetilde{\phi}_{II,\star}} \right)^{\lambda_{II,-}}\left( 1 + \frac{\delta \pi_{\star}}{\pi_{\star} + \lambda_{II,+}\widetilde{\phi}_{II,\star}} \right)^{-\lambda_{II,+}} \\ 
         &\qquad = \left( 1 + \frac{\delta \pi_{\text{end}}}{\pi_{\text{end}} + \lambda_{II,+}\widetilde{\phi}_{\text{end}}} \right)^{-\lambda_{II,+}}\left( 1 + \frac{\delta \pi_{\text{end}}}{\pi_{\text{end}} + \lambda_{II,-}\widetilde{\phi}_{\text{end}}} \right)^{\lambda_{II,-}}.
    \end{aligned}
\end{equation}

For the short-wavelength modes that exit the horizon after the transition (i.e., Case C), the curvature perturbation at the end of inflation is given by
\begin{align}
    \mathcal{R}&=\delta N=\delta N_{II}=\delta (n_{\text{end}} - n_{k} )
\nonumber\\
   &= \frac{1}{\lambda_{II,\mp}}\left[\ln\left( 1 + \frac{\delta \pi_{II}-\lambda_{II,\pm}\delta\phi_{II}}{\pi_{II} + \lambda_{II,\pm}\widetilde{\phi}_{II}} \right) -\ln\left( 1 + \frac{\delta \pi_{\text{end}}}{\pi_{\text{end}} +\lambda_{II,\pm}\widetilde{\phi}_{\text{end}}} \right)\right]\,,
\end{align}
where $\delta\pi_{\rm end}$ is implicitly determined by
\begin{equation}\label{delta pi_end2}
    \begin{aligned}
                 &\left( 1 + \frac{\delta \pi_{II}+\lambda_{II,-}\delta\phi_{II}}{\pi_{II} + \lambda_{II,-}\widetilde{\phi}_{II}} \right)^{\lambda_{II,-}}\left( 1 + \frac{\delta \pi_{II}+\lambda_{II,+}\delta\phi_{II}}{\pi_{II} + \lambda_{II,+}\widetilde{\phi}_{II}} \right)^{-\lambda_{II,+}} \\ 
         &\qquad = \left( 1 + \frac{\delta \pi_{\text{end}}}{\pi_{\text{end}} + \lambda_{II,+}\widetilde{\phi}_{\text{end}}} \right)^{-\lambda_{II,+}}\left( 1 + \frac{\delta \pi_{\text{end}}}{\pi_{\text{end}} + \lambda_{II,-}\widetilde{\phi}_{\text{end}}} \right)^{\lambda_{II,-}}.
    \end{aligned}
\end{equation}


Below we more closely examine the non-Gaussian nature of the curvature perturbation for Case A, Case B and Case C separately.

\subsubsection*{Case A: $k\ll k_\star$}

For the modes that exit the horizon well before the transition, the field fluctuation already settles down to the attractor phase when the transition occurs. Then, as discussed in \citep{Pi:2022ysn}, if the inflaton is already in the attractor phase at $\phi=\phi_{\star}$ as well as at the end of inflation $\phi=\phi_{\rm end}$, Eqs.~\eqref{perturbation of logarithmic duality} and \eqref{delta pi_end} can be simplified as
\begin{align}\label{perturbation of duality within slow roll attractor approximation}
    1 + \frac{\delta\pi_{\star}}{\pi_{\star} + \lambda_{I,-}\widetilde{\phi}_{I,\star}} &\simeq \left( 1 + \frac{\delta \pi_{I} + \lambda_{I,-}\delta\phi_{I}}{\pi_{I} + \lambda_{I,-}\widetilde{\phi}_{I}} \right)
\left( 1 + \frac{\delta \pi_{I} + \lambda_{I,+}\delta\phi_{I}}{\pi_{I} + \lambda_{I,+}\widetilde{\phi}_{I}} \right)^{-\frac{\lambda_{I,+}}{\lambda_{I,-}}}, \\
    1 + \frac{\delta\pi_{\rm end}}{\pi_{\rm end} + \lambda_{II,-}\widetilde{\phi}_{II,\rm end}} &\simeq \left( 1 + \frac{\delta \pi_{\star}}{\pi_{\star} + \lambda_{II,-}\widetilde{\phi}_{II,\star}} \right)
\left( 1 + \frac{\delta \pi_{\star}}{\pi_{\star} +\lambda_{II,+}\widetilde{\phi}_{II,\star}} \right)^{-\frac{\lambda_{II,+}}{\lambda_{II,-}}}.\label{perturbation of duality within slow roll attractor approximation 2}
\end{align}
Note that $\delta\pi_\star\to0$ and $\delta_{\rm end}\to0$ in the attractor limit. But the left-hand sides of the above equations are non-trivial because $\pi_{\star} + \lambda_{I,-}\widetilde{\phi}_{I,\star}\to0$ and $\pi_{\rm end} + \lambda_{II,-}\widetilde{\phi}_{II,\rm end}\to0$.

Substituting the above into \eqref{general R for long wavelength mode}, we obtain
\begin{align}\label{super long wavelength mode}
       \mathcal{R} &= \frac{1}{\lambda_{I,-}}\ln\left[ 1 + \frac{\delta \pi_{I} + \lambda_{I,+}\delta\phi_{I}}{\pi_{I} + \lambda_{I,+}\widetilde{\phi}_{I}} \right] +\frac{1}{\lambda_{II,-}}\ln\left[ 1 + \frac{\delta \pi_{\star}}{\pi_{\star} + \lambda_{II,+}\widetilde{\phi}_{II,\star}} \right].
\end{align}
Furthermore, under the assumption that the background is in the attractor phase initially at the first stage, we have $\pi_I+\lambda_{I,+}\widetilde{\phi}_I\simeq(\lambda_{I,+}-\lambda_{I,-})\widetilde{\phi}_I$ and $\lambda_{II,+}\widetilde{\phi}_{II,\star} + \pi_\star \simeq \lambda_{II,+}\widetilde{\phi}_{II,\star}-\lambda_{I,-}\widetilde{\phi}_{I,\star}$. Thus
except for an extremely fine-tuned case $\lambda_{II,+}\widetilde{\phi}_{II,\star}-\lambda_{I,-}\widetilde{\phi}_{I,\star}=0$,
the contribution of the second logarithmic term is negligible. 
In the end, the curvature perturbation for the super-long-wavelength modes is given by
\begin{equation}
    \mathcal{R} \simeq \frac{1}{\lambda_{I,-}}\ln\left[ 1 + \frac{\delta \pi_{I} +\lambda_{I,+}\delta\phi_{I}}{\pi_{I} + \lambda_{I,+}\widetilde{\phi}_{I}} \right] \quad\text{for}~k\ll k_\star\,. 
    \label{casea}
\end{equation}
In the limit $|\lambda_{I,-}|\ll1$, this reduces to the standard slow-roll result.

\subsubsection*{Case B: $k\lesssim k_\star$}

The modes that exit the Hubble horizon shortly before the transition point show highly non-trivial behavior.
It is not immediately apparent that the general expression \eqref{general R for long wavelength mode} conceals some non-trivial features for those modes. 
The non-trivial scale dependence of $\delta \pi_{\star}(\delta\phi,\delta\pi;\phi,\pi)$ is implicitly embedded in \eqref{logarithmic duality}. 

To analytically clarify this non-trivial dependence, we use the approximation that $-\lambda_{I,-}(n-n_\star)\ll1$. 
This enables us to approximate $e^{-\lambda_{I,-}(n-n_{\star})}\approx1$. Then from \eqref{eq:phi solution} and its $n$ derivative ($\pi=\dd\phi/\dd n$), we have
\begin{align}\label{semi-USR realation}
    \lambda_{I,+}\widetilde{\phi}_{I} + \pi_{I} \simeq
    \lambda_{X,+}\widetilde{\phi}_{I,\star}+ \pi_{\star}\,.
\end{align}
In the limit $\eta_{I}\to0$, we have $\lambda_{I,+} \to3$, this reduces to the familiar relation for an ultra-slow-roll trajectory $3\phi + \pi = 3\phi_{\star}+\pi_\star$. 
Substituting the perturbed \eqref{semi-USR realation} and \eqref{perturbation of duality within slow roll attractor approximation 2} into \eqref{general R for long wavelength mode}, we find $\delta N_I\approx0$ and obtain
\begin{equation}
    \label{near phi_* modes 1}
    \mathcal{R} \simeq \frac{1}{\lambda_{II,-}}\ln\left[ 1 + \frac{ \delta \pi_\star}{\pi_{\star} + \lambda_{II,+}\widetilde{\phi}_{II,\star}} \right]
    \simeq \frac{1}{\lambda_{II,-}}\ln\left[ 1 + \frac{\lambda_{I,+}\delta\phi_{I} + \delta \pi_{I}}{\pi_{\star} + \lambda_{II,+}\widetilde{\phi}_{II,\star}} \right] \quad \text{for}~k\lesssim k_\star\,.
\end{equation}
Thus, the main contribution is from the off-attractor behavior at the beginning of the second stage. There exists no USR term in the $\delta N$ result, although \eqref{semi-USR realation} gives very similar trajectories in phase space. 

We mention that the formula \eqref{near phi_* modes 1} differs from those reported in \citep{Cai:2022erk, Cai:2021zsp}. Specifically, the first term in (2.34) of \cite{Cai:2022erk} and in (6) of \cite{Cai:2021zsp} should be absent. This slightly modifies the remaining equations but not in an essential way as the main focus of these papers is on the effect of the upward step that leads to a cutoff in the probability distribution function of the curvature perturbation.

\subsubsection*{Case C: $k > k_\star$}
For the modes that exit the horizon after the transition, the calculations are simpler, as we only need to compute $ \delta N_{II}=\delta(n_{\rm end}-n)$. We have
\begin{equation}
\begin{aligned}\label{near phi_* modes 2}
    \mathcal{R} &= \frac{1}{\lambda_{II,-}}\ln\left[ 1 + \frac{\lambda_{II,+}\delta\phi_{II} + \delta \pi_{II}}{\pi_{II} - \lambda_{II,+}\widetilde{\phi}_{II}} \right] - \frac{1}{\lambda_{II,-}}\ln\left[ 1 + \frac{\delta \pi_{\text{end}}}{\pi_{\text{end}} - \lambda_{II,+}\widetilde{\phi}_{\text{end}}} \right] \\
    &\simeq \frac{1}{\lambda_{II,-}}\ln\left[ 1 + \frac{\lambda_{II,+}\delta\phi_{II} + \delta \pi_{II}}{\pi_{II}  + \lambda_{II,+}\widetilde{\phi}_{II}} \right] \quad \text{for}~k > k_\star\,,
\end{aligned}
\end{equation}
where again we assumed that the inflaton is already in the attractor phase at the end of inflation.
\vspace{5mm}

We now have fully nonlinear $\delta N$ results for the curvature perturbation $\mathcal{R}$ in our piecewise potential model, $\mathcal{R} = \delta N(\delta\phi, \delta\pi; \phi, \pi)$ where
$\delta\phi$ and $\delta\pi$ are normally Gaussian random variables, but $\delta N$ is a nonlinear function of $(\delta\phi,\delta\pi)$.
Although there exists $\delta \pi$-dependence in $\delta N(\delta\phi, \delta\pi)$, the variance of $\delta \pi$ is negligible compared to that of $\delta\phi$ in the combination $\delta\pi+\lambda_{+}\delta\phi$ if $|\eta|\ll1$. 
In such cases, we can further simplify the result by setting $\mathcal{R} = \delta N(\delta\phi, \delta\pi = 0)$. 
It is easy to check that the modes $k\ll k_\star$ and $k\gg k_\star$ follow the standard slow-roll result. The case given by \eqref{near phi_* modes 1} for the modes $k\lesssim k_\star$ provides an interesting example in which the curvature perturbation is not determined by $\eta_I$ at the first stage but by $\eta_{II}$ at the second stage.
Ignoring $\delta\pi_{I}$ we have
\begin{equation}
\label{near phi_* modes simpified}
\mathcal{R} \simeq \frac{1}{\lambda_{II,-}}
    \ln\left[ 1 + \frac{(3-\eta_I)\delta\phi_{I} }{\pi_{\star} + \lambda_{II,+}\sqrt{2\epsilon_{II}}/\eta_{II}} \right]\,,
\end{equation}
where we have assumed $|\eta_I|\ll1$ for simplicity, but $\eta_{II}$ is not necessarily small. Using $\pi_\star$ given by \eqref{pistar} and the fact $|\eta_I|\ll1$, we have $\pi_\star\simeq-\sqrt{2\epsilon_I}$. Hence the above equation can be concisely expressed as
\begin{align}\label{nonlinear log R_G}
    \mathcal{R} = \frac{1}{\gamma} \ln \left[1 + \gamma\,\mathcal{R}_{\rm G}\right]\,;\quad
    \mathcal{R}_{\rm G} \equiv 
     \frac{(3-\eta_I)\delta\phi_{I} }{
     \lambda_{II,-}( \lambda_{II,+}\sqrt{2\epsilon_{II}}/\eta_{II} - \lambda_{I,-}\sqrt{2\epsilon_I} /\eta_{I})} 
\end{align}
where $\gamma=\lambda_{II,-}$ and $\mathcal{R}_{\rm G}$ is the Gaussian part of the curvature perturbation. Expanding the logarithmic function, one finds
\begin{equation}
\mathcal{R}
=\mathcal{R}_\mathrm{G}+\frac{3}{5}f_{\rm NL}\mathcal{R}_\mathrm{G}^2+\cdots\,;
 \quad f_{\rm NL}=-\frac{5}{6}\gamma\,.
\end{equation}
Since $\gamma=\lambda_{II,-}=(3-\sqrt{9-12\eta_{II}})/2$, the non-Gaussian parameter $f_{\rm NL}$ can be large if $\eta_{II}$ is large and negative.

As in the case of \eqref{near phi_* modes simpified} above, when one of the logarithmic functions dominates $\mathcal{R}$, one can always express the curvature perturbation in the form of \eqref{nonlinear log R_G}, except for the form of $\mathcal{R}_\mathrm{G}$. 
In such a case, using the probability conservation $\mathbb{P}[\mathcal{R}] \dd \mathcal{R}= \mathbb{P}[\mathcal{R}_{\rm G}] \dd \mathcal{R}_{\rm G}$, we can easily get the probability distribution function (PDF) of $\mathcal{R}$,
\begin{align}
    \mathbb{P}[\mathcal{R}]  = \frac{1}{\sqrt{2\pi}\sigma_{\mathcal{R}_{\rm G}}}\exp \left[ \gamma\, \mathcal{R} - \frac{(e^{\gamma\,\mathcal{R}}-1)^2}{2\gamma^2 \sigma^2_{\mathcal{R}_{\rm G}}} \right]\,,
\end{align}
where $\sigma^2_{\mathcal{R}_{\rm G}}$ is the variance of the Gaussian part of the PDF, $\mathbb{P}[\mathcal{R}_{\rm G}]$,
\begin{equation}
    \sigma^2_{\mathcal{R}_{\rm G}}=\int\mathcal{R}_\mathrm{G}^2\mathbb{P}[\mathcal{R}_{\rm G}]d\mathcal{R}_\mathrm{G}\,.
\end{equation}
In particular, there always appears an exponential tail in the PDF at large positive $\mathcal{R}$ if $\beta<0$ and at large negative $\mathcal{R}$ if $\beta>0$. 

\subsection{Numerical algorithm and comparison with analytical formulas}
\label{appendix: numerical delta N}
To check the robustness of our analytical approximations \eqref{super long wavelength mode}, \eqref{near phi_* modes 1}, and \eqref{near phi_* modes 2}, we numerically evaluated the exact $\delta N$ formula  \eqref{general R for long wavelength mode} for Case A, Case B and Case C. 
Here we first describe the numerical algorithm we adopted.

As mentioned in sec.~\ref{sec: The non-Gaussianity: deltaN calculation}, there are three situations where perturbation modes exit the Hubble horizon at different times. However, the consistency of the analytical approximations throughout the entire transition from the first slow-roll stage to the second has not been fully clear yet. 
In particular, for the modes crossing the horizon very near the transition, the field fluctuation determines whether the horizon crossing occurs before or after the transition. 
Thus the fluctuation significantly affects the corresponding background trajectory, leading to fully non-perturbative phenomena. 

The key to analytically determining the nonlinear formula for $\mathcal{R}(\delta \phi)$ in \eqref{general R for long wavelength mode} is to explicitly express $\delta \pi_{\star}(\phi,\pi)$ and $\delta \pi_{\rm end}(\pi_{\star})$. 
These quantities are determined by perturbing the \textit{logarithmic duality} \eqref{perturbation of logarithmic duality} or \eqref{delta pi_end}, which typically requires numerical evaluations.
The special cases in which it can be algebraically solved occur when $\lambda_{X,+}/\lambda_{X,-} = m/n$, where $m, n \in \mathbb{Z}$, $m > |n|$, and $\text{max}(m,m-n) < 4$. The details about solvable cases are given in \cite{Pi:2022ysn}. 
One option is to solve \eqref{perturbation of logarithmic duality} or \eqref{delta pi_end} numerically and substitute the result into the expression for $\mathcal{R}$. However, here we adopt a more straightforward approach.

The algorithm is outlined in the following steps:
\begin{itemize}
    \item Step 1: Write down the analytical solution of the equation of motion (EOM) \eqref{eq:Background equation in delta N} for both stages, using the initial conditions $(\phi_{\rm i}, \pi_{\rm i})$ for the first stage $X = I$ and $(\phi_{\rm i}, \pi_{\rm i})$ for the second stage $X = II$:
        \begin{align} 
            \phi_{I}(n) &= \phi_{\star} -\frac{\sqrt{2\epsilon_{I}}}{\eta_{I}} - \frac{\pi_{\rm i} +  \lambda_{I,+}(\phi_{\rm i} - \phi_{\star} + \sqrt{2\epsilon_{I}}/\eta_{I})}{\lambda_{I,-} - \lambda_{I,+}}e^{-\lambda_{I,-}(n-n_{\rm{i}})} \nonumber\\
            &\qquad+ \frac{\pi_{\rm i} + \lambda_{I,-}(\phi_{\rm i} - \phi_{\star} + \sqrt{2\epsilon_{I}}/\eta_{I})}{\lambda_{I,-} - \lambda_{I,+}}e^{-\lambda_{I,+}(n-n_{\rm{i}})}\,; \quad n < n_{\star}\,,\label{eq:phi_solution_with_initial_I}\\
            \phi_{II}(n) &= \phi_{\star} -\frac{\sqrt{2\epsilon_{II}}}{\eta_{II}} - \frac{\pi_{\rm i} + \lambda_{II,+}(\phi_{\rm i} - \phi_{\star} + \sqrt{2\epsilon_{II}}/\eta_{I})}{\lambda_{II,-} - \lambda_{II,+}}e^{-\lambda_{II,-}(n-n_{\rm{i}})} \nonumber\\
            &\qquad+ \frac{\pi_{\rm i} +  \lambda_{II,-}(\phi_{\rm i} - \phi_{\star} + \sqrt{2\epsilon_{II}}/\eta_{II})}{\lambda_{II,-} - \lambda_{II,+}}e^{-\lambda_{II,+}(n-n_{\rm{i}})}\,; \quad n > n_{\star}\,.\label{eq:phi_solution_with_initial_II}
        \end{align}
    \item Step 2: Choose an initial condition $(\phi_{\rm i}, \pi_{\rm i})$ at the first stage that ensures the background evolution is physically valid. Then, evaluate $\pi_{\star}\equiv \dd \phi_{I}(n_{\star})/\dd n$ and the e-folding number from $\phi_{\rm i}$ to $\phi_{\star}$: $N_{I} = n_{\star} - n_{\rm i}$.
    \item Step 3: Given the end of inflation $\phi_{\rm end}$, solve for the e-folding number $N_{II} = n_{\rm end} - n_{\star}$ from $\phi_{\star}$ to $\phi_{\rm end}$, using $(\phi_{\star}, \pi_{\star})$ as the initial condition for  \eqref{eq:phi_solution_with_initial_II}.
    \item Step 4: Sum the total e-folding number $N = N_I + N_{II}$ for the background solution.
    \item Step 5: Calculate the new total e-folding number by introducing a different initial condition $(\phi_{\rm i} + \delta\phi, \pi_{\rm i} + \delta \pi)$. 
    The key point here is to check whether the perturbed $\phi_{\rm i} + \delta\phi$ is greater or less than $\phi_{\star}$. 
    If $\phi_{\rm i} + \delta\phi > \phi_{\star}$, then we repeat Steps 1 to 3. 
    On the other hand, if  $\phi_{\rm i} + \delta\phi < \phi_{\star}$, we obtain the total e-folding number by solving $N = N_{II} = n_{\rm end} - n_{\rm i}$.
    \item Step 6: By repeating the above, numerically establish the dependence of the total e-folding number as a function of the initial condition, $\delta N(\phi_{\rm i}, \pi_{\rm i}; \delta\phi, \delta\pi)$.
\end{itemize}
By employing this algorithm, we can generate a contour plot of numerically computed $\delta N$ for a set of initial conditions $(\phi_{\rm i}+\delta\phi,\pi_{\rm i}+\delta\pi)$. 

\begin{figure}[htbp]
\centering
\includegraphics[width=0.8\textwidth]{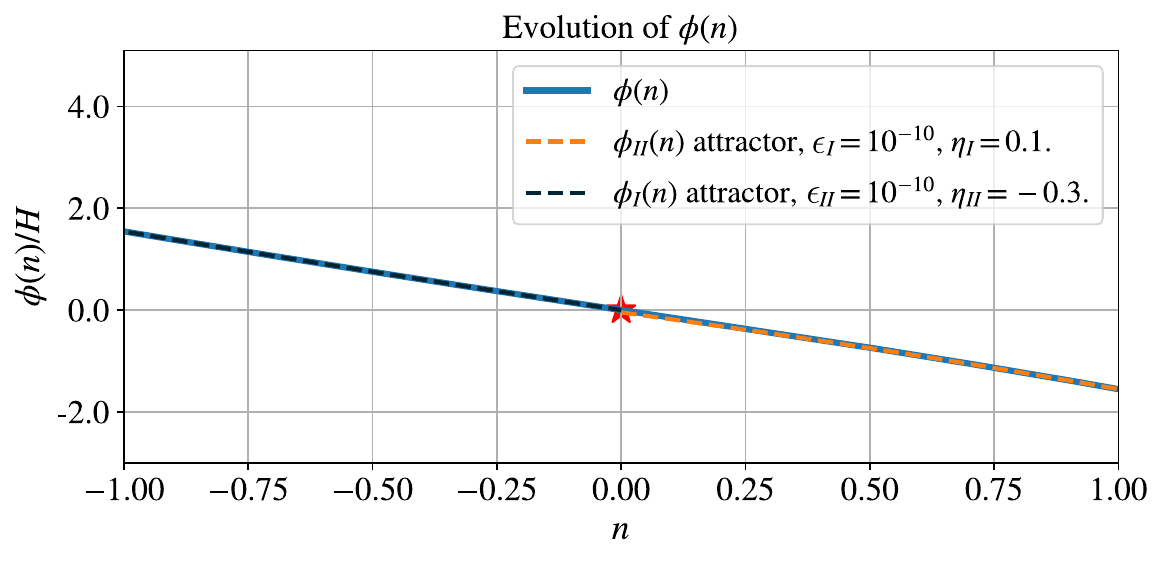}
\includegraphics[width=0.8\textwidth]{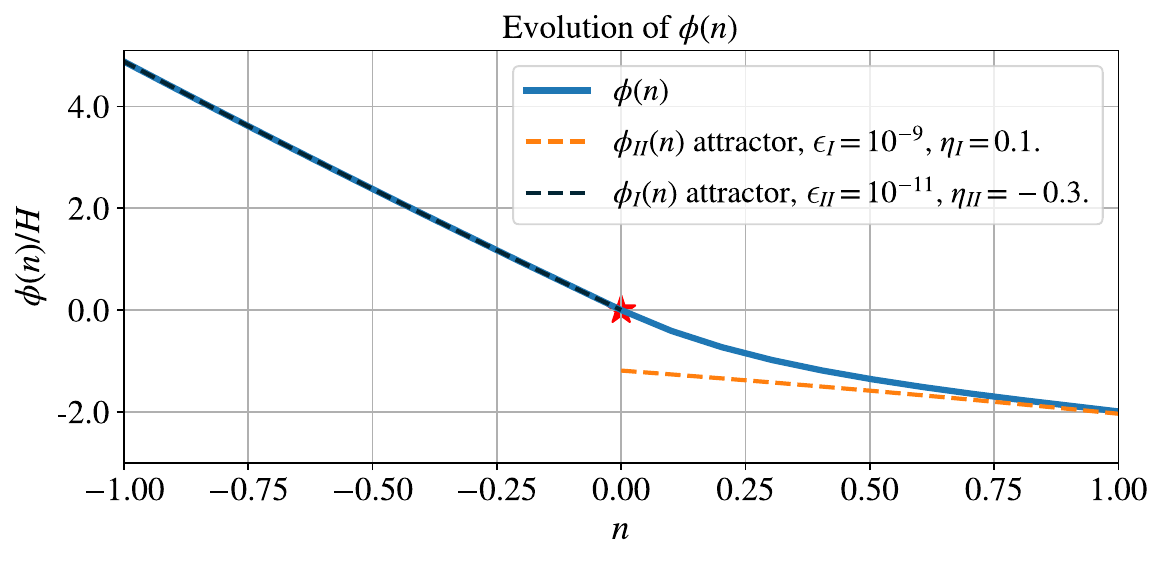}
\caption{The evolution of $\phi(n)$ is governed by Equations \eqref{eq:phi_solution_with_initial_I} and \eqref{eq:phi_solution_with_initial_II}. We select $\pi_{\star} = 1.03576\sqrt{2\epsilon_{I}}$ for the blue line. The red star marks the joint point at $(n = 0, \phi = 0)$.
}
\label{fig:background phi}
\end{figure}

We select the same first slow-roll parameter, $\epsilon_{I} = \epsilon_{II} = 10^{-10}$, to isolate and highlight the effect of the second slow-roll parameter $\eta_V$. 
With $\epsilon_{V} = 10^{-10}$, a typical inflationary energy scale of $H/M_{\rm p} = 10^{-5}$ can enhance the primordial power spectrum to $\mathcal{O} (10^{-2})$.
The background evolution of $\phi(n)$ is illustrated in Fig.~\ref{fig:background phi}.
The blue line represents the analytical results from \eqref{eq:phi_solution_with_initial_I} and \eqref{eq:phi_solution_with_initial_II}, with $\phi_{\star} = 0$ and $\pi_{\star} = 1.03576\sqrt{2\epsilon_{I}}$. The dashed orange and black lines indicate the attractor solution. Here, we set $n_{\star} = 0$.

\begin{figure}[htbp]
\centering
\includegraphics[width=0.8\textwidth]{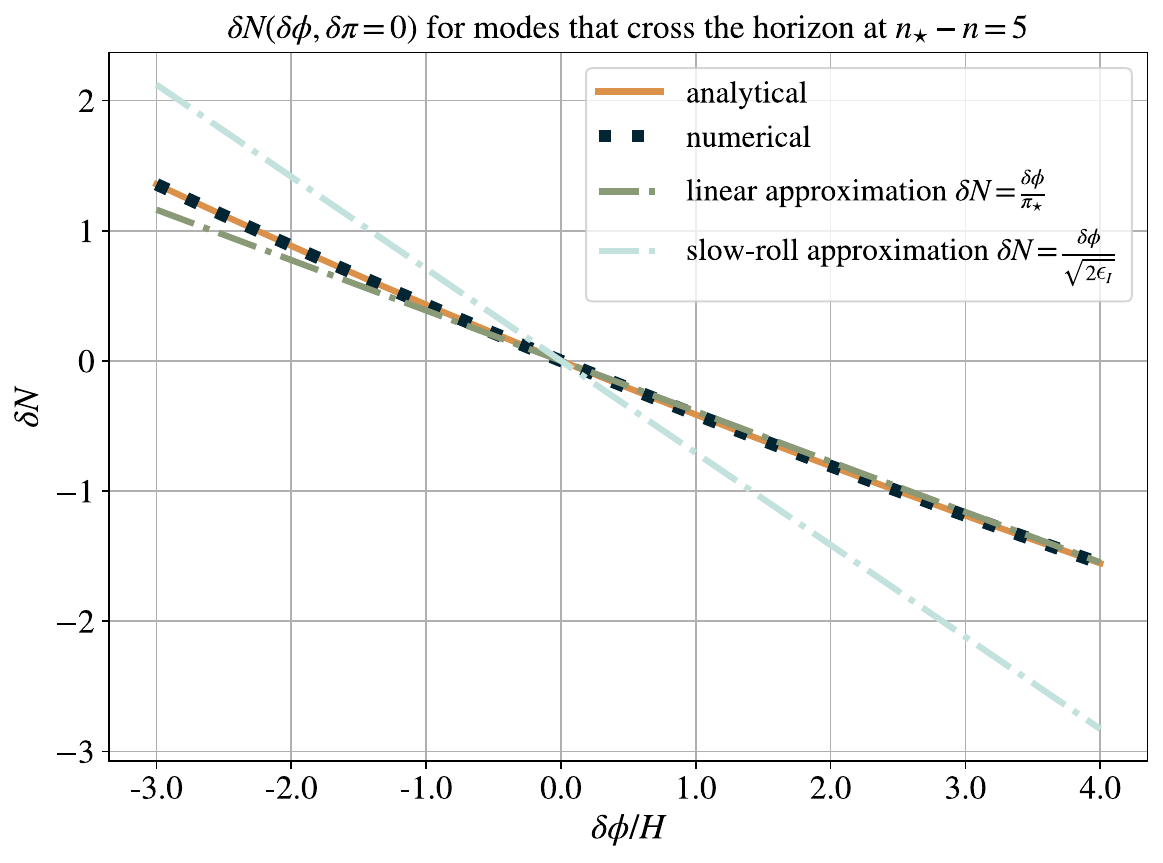}
\caption{We consider the point \((\phi_{\rm i}, \pi_{\rm i})\) on the blue line in fig.~\ref{fig:background phi}, 5 e-folds before the inflaton reaches the joint point, and evaluate the numerical \(\delta N\). The parameters are set as \(\epsilon_{I} = \epsilon_{II} = 10^{-10}\), \(\eta_{I} = 0.1\), and \(\eta_{II} = -0.3\).}
\label{fig:super long numerical deltaN}
\end{figure}

As shown in Fig.~\ref{fig:background phi}, $\phi(n)$ has already converged to the first-stage attractor solution when $n_{\rm i} - n_{\star} = -5$. We select $(\phi, \pi)$ at $n_{\rm i} - n_{\star} = -5$ to illustrate Case A; see Fig.~\ref{fig:super long numerical deltaN}. 
The orange line represents the analytical result from \eqref{casea} for Case A and matches the numerical $\delta N$ result well. 
We also observe a deviation from linearity at $\delta N \approx \mathcal{O}(1)$, which may be important for PBH formation.

\begin{figure}[htbp]
\centering
\includegraphics[width=0.8\textwidth]{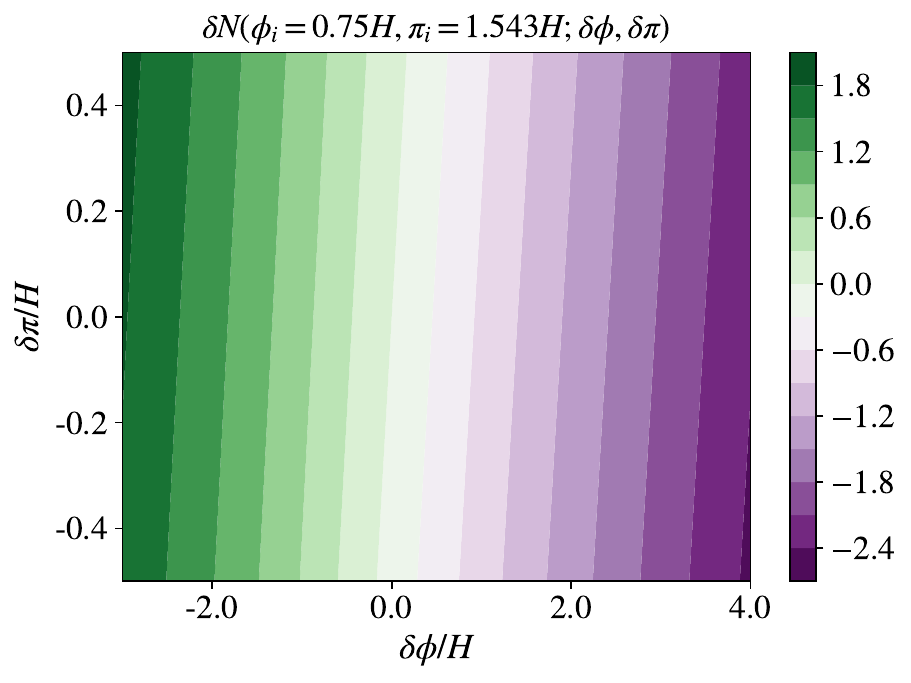}
\caption{Contour plot of $\delta N(\delta\phi,\delta \pi)$ for modes crossing the Hubble horizon 5 e-folds before the joint point. The parameters are set as $\epsilon_{I} = \epsilon_{II} = 10^{-10}$, $\eta_{I} = 0.1$, and $\eta_{II} = -0.3$. We set $\delta \pi = 0$ to obtain the numerical dashed line in the top panel of fig.~\ref{fig:numerical deltaN and PDF}.}
\label{fig:delta N contour}
\end{figure}

\begin{figure}[htbp]
\centering
\includegraphics[width=0.95\textwidth]
{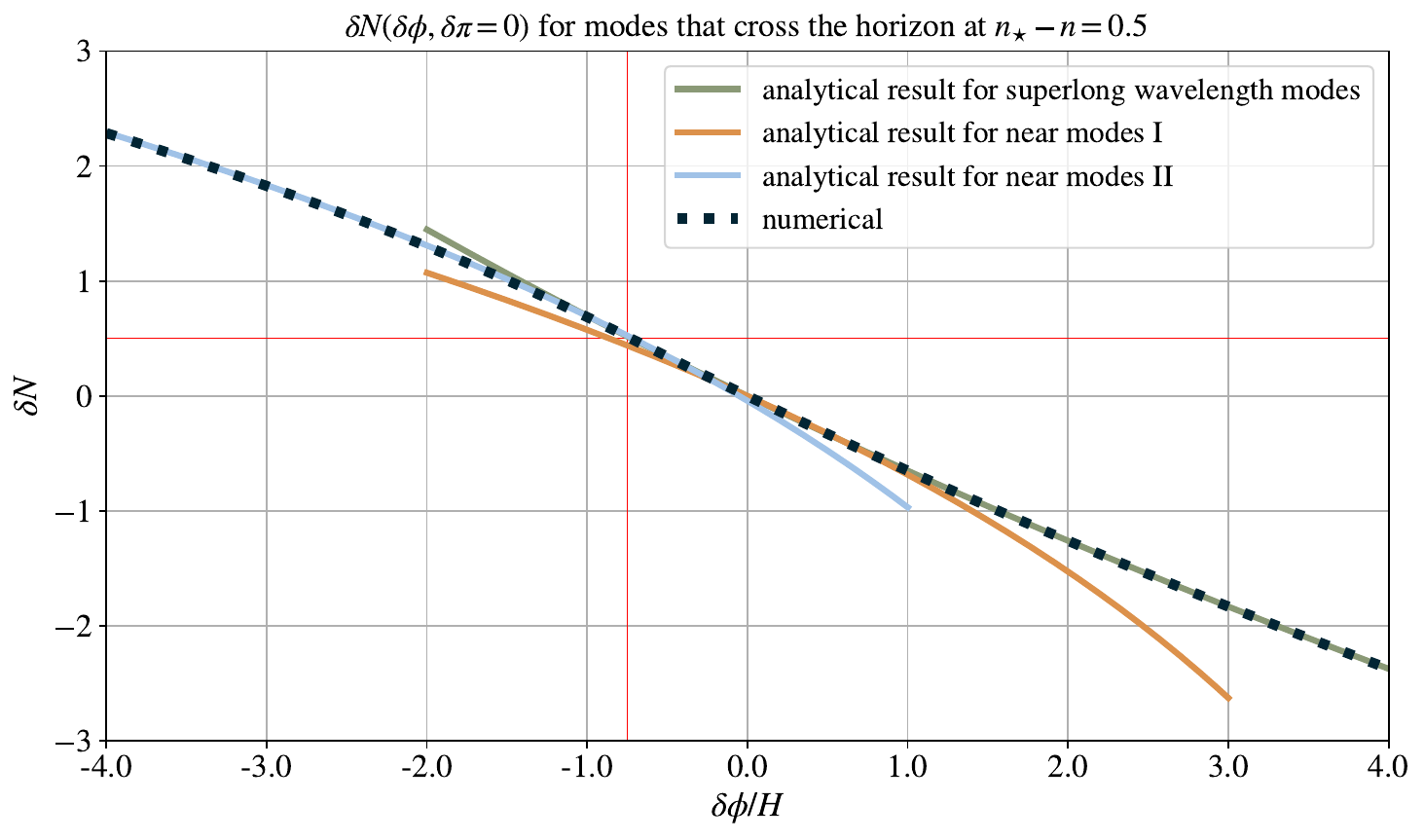}
\includegraphics[width=0.95\textwidth]{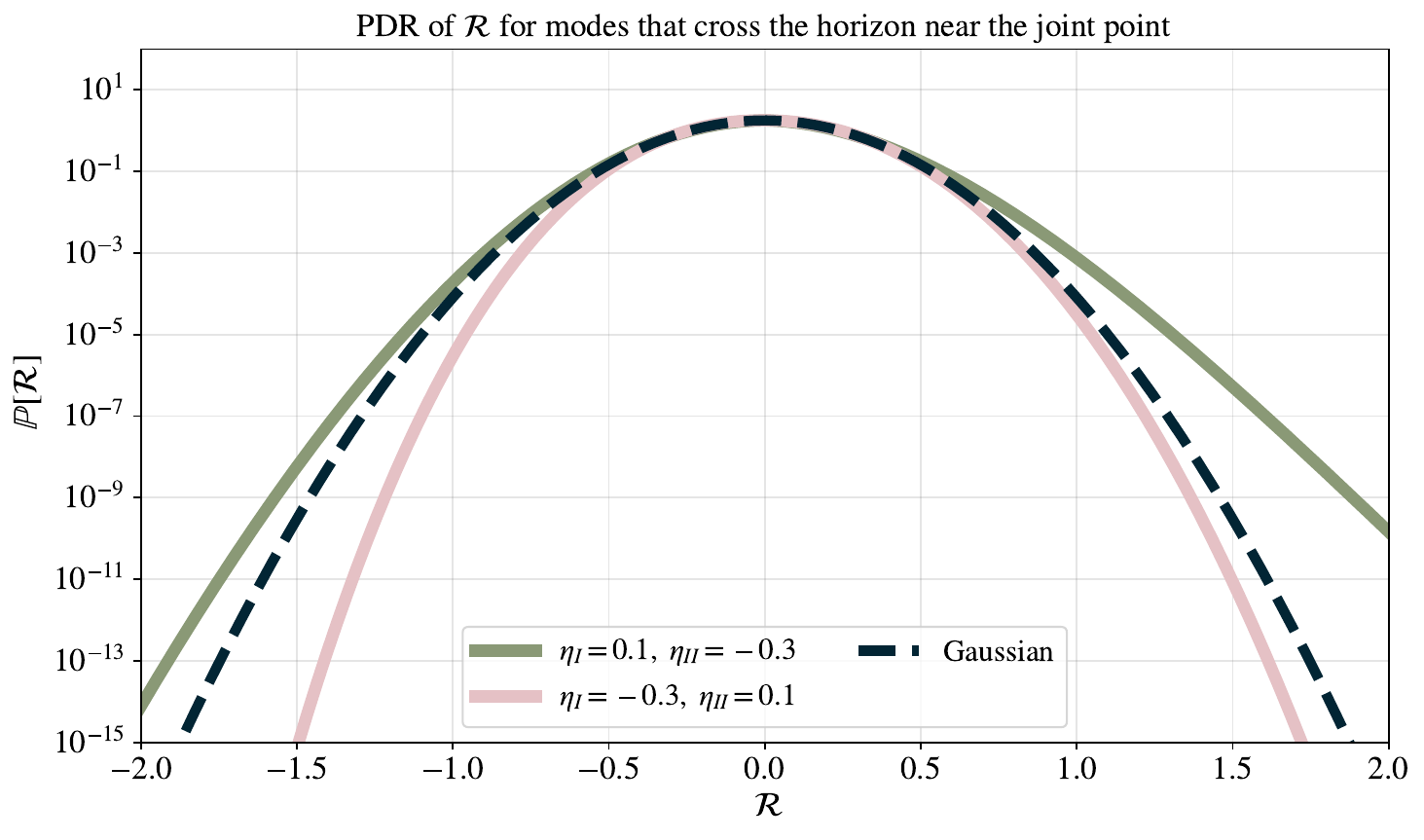}
\caption{\textbf{Top panel:} We demonstrate the nonlinear behavior of the curvature perturbation \eqref{general R for long wavelength mode} for modes crossing the horizon near the joint point $\phi = \phi_{\star}$. We choose parameters as $\epsilon_{I} = \epsilon_{II} = 10^{-10}$, $\eta_{I} = 0.1$, and $\eta_{II} = -0.3$ for demonstration. The black dots denote numerical $\delta N$ results. The modes crossing the horizon before and after the joint point are represented on the right and left sides of the vertical thin red solid line, respectively. The blue line represents the approximate result of \eqref{near phi_* modes 2}, the orange line represents \eqref{near phi_* modes 1}, and the green line represents \eqref{super long wavelength mode}.
\textbf{Bottom panel:} The nonlinearity of $\mathcal{R}(\delta\phi)$ causes $\mathbb{P}[\mathcal{R}]$ to deviate from a Gaussian distribution. The PDF of the green line is given by the fully nonlinear relation demonstrated in the top panel. Changing the parameters leads to the pink one. There are always exponential tails in the PDF; however, the exponential tail on the positive $\mathcal{R}$ side is determined by $\eta_{II}$, while the exponential tail on the negative side is determined by $\eta_{I}$.}
\label{fig:numerical deltaN and PDF}
\end{figure}

To study the modes of most interest, Cases B and C,  we choose $n_{\rm i}-n_{\star} = 0.5$ for a complete numerical demonstration, as shown in Fig.~\ref{fig:delta N contour}. 
At $n=n_{\rm i}$, the background solution yields $\phi_{\rm i} = 0.75 H$ and $\pi_{\rm i } = 1.543 H$ (represented by the solid blue line in Fig.~\ref{fig:background phi}). This makes it easy for the quantum fluctuation of $\delta \phi$ to shift the initial data, $(\phi_{\rm i}+\delta\phi,\pi_{\rm i}+\delta\pi)$, beyond the transition point. 
This mechanism highlights the non-perturbative phenomena observed in our model. 
As shown by the contour lines of $\delta N$ in Fig.~\ref{fig:delta N contour}, the intervals are non-uniform along the $\delta \phi$ direction, which directly reflects the nonlinearity observed in the top panel of Fig.~\ref{fig:numerical deltaN and PDF}. 

It is worth noting that $\delta \pi$ also contributes to $\delta N$, as indicated by the slope of the contour lines. We typically ignore the contribution of $\delta \pi$ because $\delta N$ is valid when the modes are superhorizon, i.e., several e-folds after horizon crossing. The $\delta \pi$ term decays exponentially and can therefore be ignored.

In the case when there is a sharp transition feature like in the current model, another noteworthy feature appears in the PDF. Consider the mode that exits the Hubble horizon at the transition point, $k=k_\star$. As in our model, let us assume that the background inflaton rolls down toward smaller values, $d\phi/dn<0$. Then the field exits the horizon {\it before} the transition if $\delta\phi>0$, while it exits the horizon {\it after} the transition if $\delta\phi<0$. 
This implies that depending on the sign of $\delta\phi$, one has to use the different formulas of  $\mathcal{R}$. For example, \eqref{near phi_* modes 1} for $\delta\phi<0$ and \eqref{near phi_* modes 2} for $\delta\phi>0$. 
This further implies that the formula to be used depends on the sign and the magnitude of $\delta\phi$. 
For example, for a mode that leaves the horizon slightly before the transition, $k=k_\star-\Delta k$ ($\Delta k>0$), \eqref{near phi_* modes 1} should be used for $\delta\phi>-\epsilon$ ($\epsilon>0$) while \eqref{near phi_* modes 2} should be used for $\delta\phi<-\epsilon$. 
We illustrate this interesting nonlinear effect in fig.~\ref{fig:numerical deltaN and PDF}, which are obtained numerically by using the complete expression of $\mathcal{R}$,  \eqref{general R for long wavelength mode}. 
 
Here it is important to recall that the evaluation of the curvature perturbation by the $\delta N$ formula does not depend on the time at which one evaluates the field fluctuations as long as the scale is superhorizon and the evolution is classical. Thus the above argument is independent of the precise definition of the horizon crossing time. 
In fact, one can evaluate $\delta N$ at any time after horizon crossing. We will come back to this point in the next subsection.

\subsection{Nonlinear evolution of superhorizon modes}
In this subsection, we discuss the conservation of the comoving curvature $\mathcal R$ in the nonlinear regime.
Although it is already proved in \cite{Lyth:2004gb} generically, we find it useful to consider our case explicitly.
 
Let us focus on the super-long wavelength (SLWL) modes, say, on the scales measured by the cosmic microwave background \cite{Planck:2018jri}. 
We know from the observational data that the slow-roll inflation is valid on those scales. The SLWL modes are frozen, and the spatial gradients can be ignored soon after horizon crossing. Thus, we can evaluate $\delta N$ soon after horizon crossing.
Namely, we use \eqref{casea} with $\delta\phi$ inferred from linear perturbation theory.

Now, let us evaluate $\delta N$ of the SLWL modes after the epoch of transition, $n>n_\star$. Then we should use \eqref{near phi_* modes 2}. But the formula looks completely different from \eqref{casea}. So if one would naively solve for $\delta\phi$ by using the linear perturbation equation, one would obtain a completely wrong answer.
This seems a contradiction as one would expect that $\delta\phi$ would follow the linear evolution equation since the scalar field evolution is linear already at the background level.



The answer is that the evolution of $\delta\phi$ is actually highly nonlinear on superhorizon scales when there is a sharp feature in the potential.
Let us explain the origin of the nonlinearity. To do so, we first write down the solution of $\phi$ by neglecting the spatial-gradient terms at two different times $n=n_1$ and $n_2$, where $n_1$ belongs to the first stage and $n_2$ to the second stage. Assuming that the system is in the attractor regime both at $n=n_1$ and $n_2$, 
we have 
\begin{equation}
\begin{aligned}
    &\widetilde{\phi_I}(n_1)=C_{I,-}\exp(-\lambda_{I,-}(n_1-n_\star)),\\
    &\widetilde{\phi_{II}}(n_2)=C_{II,-}\exp(-\lambda_{II,-}(n_2-n_\star)),\\
\end{aligned}
\end{equation}
where the time when $\phi=\phi_\star$ is given by
\begin{align}
      n_\star=n_1+\frac{1}{\lambda_{I,-}}\ln\left(\frac{\widetilde{\phi_I}_\star}{\widetilde{\phi_I}(n_1)}\right)\,.    
\end{align}
As in the case of $\phi_\star$ being the end of inflation, $n_\star$ is a nonlinear function of the initial condition $\phi(n_1)$. 
Inserting the expression of $n_\star$ into $\widetilde{\phi_{II}}(n_2)$, we clearly see that the relation between $\widetilde{\phi_{II}}(n_2) $ and  $\widetilde{\phi_I}(n_1)$ is nonlinear.
Setting $\widetilde{\phi_I}(n_1)\to \widetilde{\phi_I}(n_1)+\delta\phi(n_1)$ and
$\widetilde{\phi_{II}}(n_2)\to \widetilde{\phi_{II}}(n_2)+\delta\phi(n_2)$, we obtain $\delta\phi(n_2)$ as a highly nonlinear function of $\delta\phi(n_1)$, 
\begin{equation}
	\delta\phi(n_2)    =\widetilde{\phi_{II}}(n_2) \Biggl\{
	\left(\frac{\widetilde{\phi_I}_\star}{\widetilde{\phi_I}(n_1)}\right)^{\lambda_{II,-}/\lambda_{I,-}}
-	\left(\frac{\widetilde{\phi_I}_\star}{\widetilde{\phi_I}(n_1)+\delta\phi(n_1)}\right)^{\lambda_{II,-}/\lambda_{I,-}}
	\Bigg\}.
\end{equation}
Thus, assuming $\delta\phi$ is Gaussian at horizon crossing does not necessarily mean that it stays Gaussian afterward. 
As expected, if we take this nonlinearity into account in the evolution of $\delta\phi$, $\mathcal{R}$ on superhorizon scales for the SLWL modes remain conserved.

In passing, we mention its possible implication to the recent dispute over the loop effects on the curvature perturbation on superhorizon scales~\cite{Kristiano:2022maq,Kristiano:2023scm,Kristiano:2024vst,Inomata:2024lud,Riotto:2023hoz,Choudhury:2023vuj,Riotto:2023gpm,Firouzjahi:2023aum,Franciolini:2023agm,Tasinato:2023ukp,Cheng:2023ikq,Fumagalli:2023hpa,Maity:2023qzw,Tada:2023rgp,Firouzjahi:2023bkt,Davies:2023hhn,Iacconi:2023ggt,Saburov:2024und,Kawaguchi:2024rsv,Ballesteros:2024zdp}.
As the field perturbation $\delta\phi$ in our computations corresponds to the one on flat slicing in perturbation theory, it implies that the conservation of the curvature perturbation cannot be guaranteed even at the classical level unless the nonlinear corrections to $\delta\phi$ are properly taken into account, if the perturbation is evolved in the flat slicing.

\section{Primordial black holes from non-Gaussian tails}
\label{sec:PBHs with non-Gaussian tail}

In this section, we consider the effect of non-Gaussianity on the abundance of primordial black holes (PBHs) in our piecewise quadratic potential model.
Since detailed discussions on the formation criteria are beyond the scope of the present paper, for the sake of simplicity, we adopt the Press-Schechter formalism in which the threshold for the PBH formation is given by a critical amplitude of the density contract smoothed over a given comoving scale at the time of the horizon re-entry after inflation. 
Further, we ignore the nonlinearity arising from the Einstein equation and set the threshold in the framework of linear perturbation theory.

In linear theory, the density contrast $\delta\equiv\rho/\bar{\rho}-1$ in the comoving slicing is related to the comoving curvature perturbation as
\begin{equation}
 \delta= -\frac{2}{3}\frac{3+3\omega}{5+3\omega}
 \left(\frac{1}{aH}\right)^2\nabla^{2}\mathcal{R}\,,   
 \label{lineardelta}
\end{equation}
where $w=1/3$ during the radiation-dominated era, which we assume in the following. 
If we consider a comoving scale $R_s$, \eqref{lineardelta} at the time of horizon re-entry, $R_s=aH$, gives
\begin{equation}
    \delta\approx\frac{4}{9}\mathcal{R}\,,
\end{equation}
where we have assumed that the smoothing on the scale $R_s$ effectively singles out the amplitude of the mode $k=1/R_s$ in Fourier space.
Then it follows that the threshold of $\delta$ is simply related to that of $\mathcal{R}$ by $\delta_{\rm th}=9/4\mathcal{R}_{\rm th}$.
Namely, we have the linear relation between the thresholds of the density contrast threshold and the comoving curvature perturbation.
Furthermore, assuming that we evaluate $\delta N$ when the field fluctuation is still Gaussian, 
the Gaussian part of the curvature perturbation is linearly related to the field fluctuation by $\mathcal{R}_{\rm G}=-\delta\phi/(d\phi/dn)=\delta\phi/\pi$ as in the standard slow-roll case. 

Under the above approximations, the abundance of PBHs at the time of formation on the comoving scale $R_s$ is determined by the threshold $\mathcal{R}_{\rm th}(R_s)$ as
\begin{align}
    \beta_{\rm PBH}(R_s) &\equiv \frac{\rho_{\rm PBH}(R_s)}{\rho_{\rm tot}} 
    = \int^{+\infty}_{\mathcal{R}_{\rm th}(R_s)}\mathbb{P}[\mathcal{R}]\dd \mathcal{R}
    =\int^{+\infty}_{\mathcal{R}_{\rm G,th}(R_s)}\mathbb{P}[\mathcal{R}_{\rm G}]\dd \mathcal{R}_{\rm G}
   \nonumber \\
    &= \int^{+\infty}_{\delta\phi_{\rm th}(R_s)}\frac{1}{\sqrt{2\pi}\sigma_{\delta\phi}(R_s)}\exp\left[-\frac{\delta\phi^2}{2\sigma^2_{\delta\phi}(R_s)}\right]\dd \delta\phi\,
    \nonumber\\
   & =\mathrm{erfc}\left(\frac{\delta\phi_{\rm th}(R_s)}{\sqrt{2}\sigma_{\delta\phi}(R_s)}\right)\,,
\end{align}
where the probability conservation $\mathbb{P}[\mathcal{R}] \dd \mathcal{R}= \mathbb{P}[\mathcal{R}_{\rm G}] \dd \mathcal{R}_{\rm G}$ and the relation between the Gaussian variables $\mathcal{R}_{\rm G}=\delta\phi/\pi$ have been used, where the Gaussianity of $\delta\phi$ is assumed. 
Denoting $\mathcal{R}$ as a function of $\mathcal{R}_{\rm G}$ by $\mathcal{R}=F(\mathcal{R}_{\rm G})$, the threshold for each variable is related as $\delta\phi_{\rm th}=\pi\mathcal{R}_{\rm G,th}=\pi F^{-1}(\mathcal{R}_{\rm th})$, where $F^{-1}$ is the inverse function of $F$.


Now, we discuss the dependence of the PBH abundance on the degree of non-Gaussianity. As an interesting specific example for the function $F$, let us consider the curvature perturbation for the modes that cross the horizon just before the transition, given by \eqref{nonlinear log R_G}. 
The threshold for $\delta\phi$ is determined by that for $\mathcal{R}$ as
\begin{align}
\delta\phi_{\mathrm{th}}=\frac{1-e^{\lambda_{II,-}\mathcal{R}_{\mathrm{th}}}}{3}
\left(\sqrt{2\epsilon_I}-\frac{\lambda_{II,+}}{{\eta_{II}}}\sqrt{2\epsilon_{II}}\right)\,,
    \label{deltaphith}
\end{align}
where we have taken the limit $\eta_I\to0$ as its effect is unimportant.
In comparison, if we would ignore the non-Gaussianity, the threshold for $\delta\phi$ would become
\begin{align}
\delta\phi_{\mathrm{th}}^{\mathrm{G}}=-\frac{\lambda_{II,-}}{3}\mathcal{R}_{\rm th}
\left(\sqrt{2\epsilon_I}-\frac{\lambda_{II,+}}{{\eta_{II}}}\sqrt{2\epsilon_{II}}\right),
    \label{deltaphithG}
\end{align}
Here, combining the above two equations, we find the effect of non-Gaussianity on the threshold value of $\delta\phi$ as
\begin{align}
    \delta\phi_{\mathrm{th}}=\frac{1-e^{\lambda_{II.-}\mathcal{R}_{\rm th}}}{-\lambda_{II,-}\mathcal{R}_{\rm th}} \delta\phi_{\mathrm{th}}^{\mathrm{G}}
    =\left(1+\frac{\eta_{II}\mathcal{R}_{\rm th}}{2}+\frac{(\eta_{II}\mathcal{R}_{\rm th})^2}{3}+\mathcal{O}(\eta_{II}^3)\right)\delta\phi_{\mathrm{th}}^{\mathrm{G}}\,,
    \label{th}
\end{align}
for a given threshold value of $\mathcal{R}_{\rm th}$, 
where the second equality is for $|\eta_{II}|\ll1$.

\begin{figure}[htbp]
\centering
\includegraphics[width=0.8\textwidth]{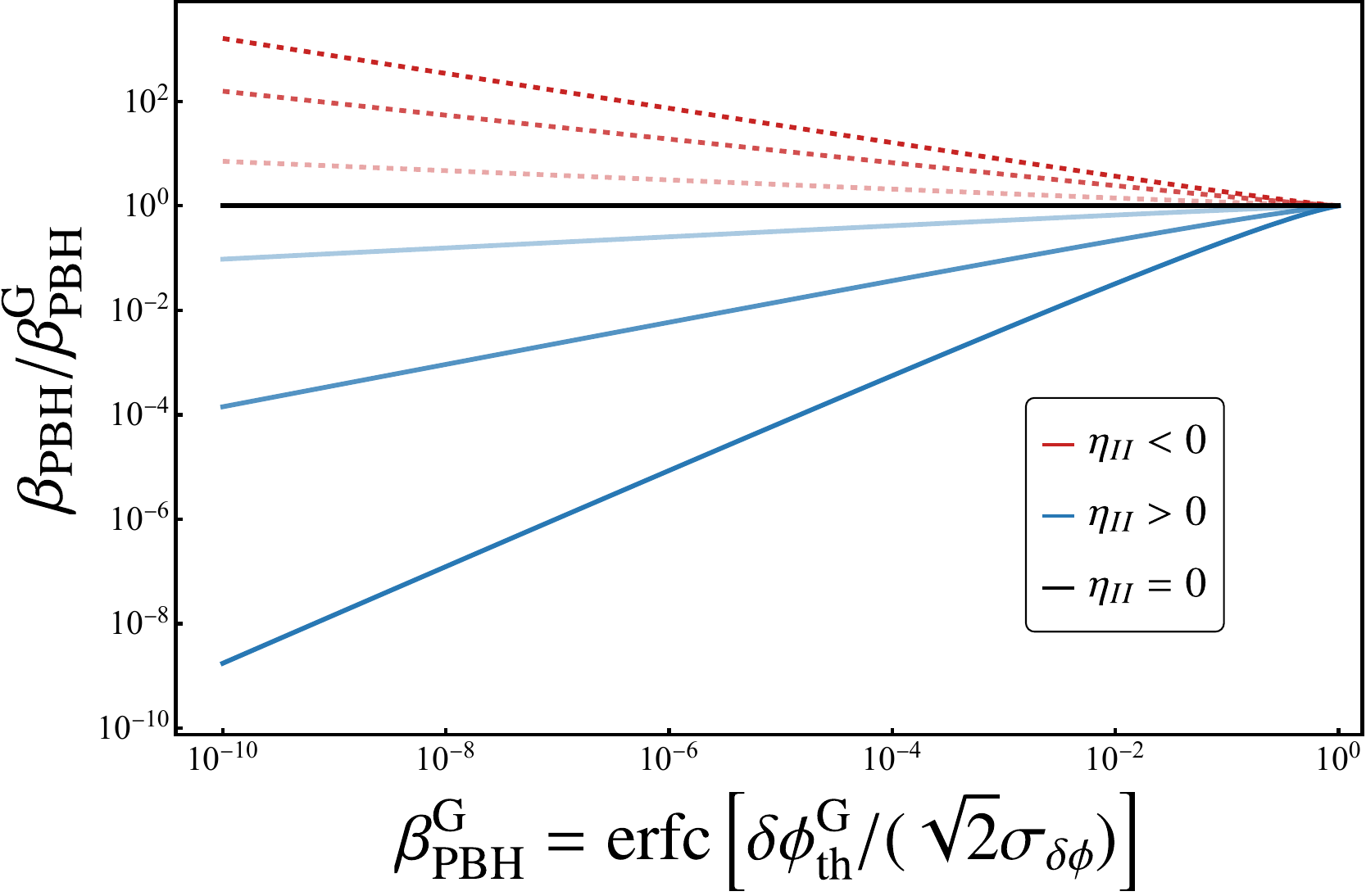}
\vspace{8mm}

\centering
\includegraphics[width=0.8\textwidth]{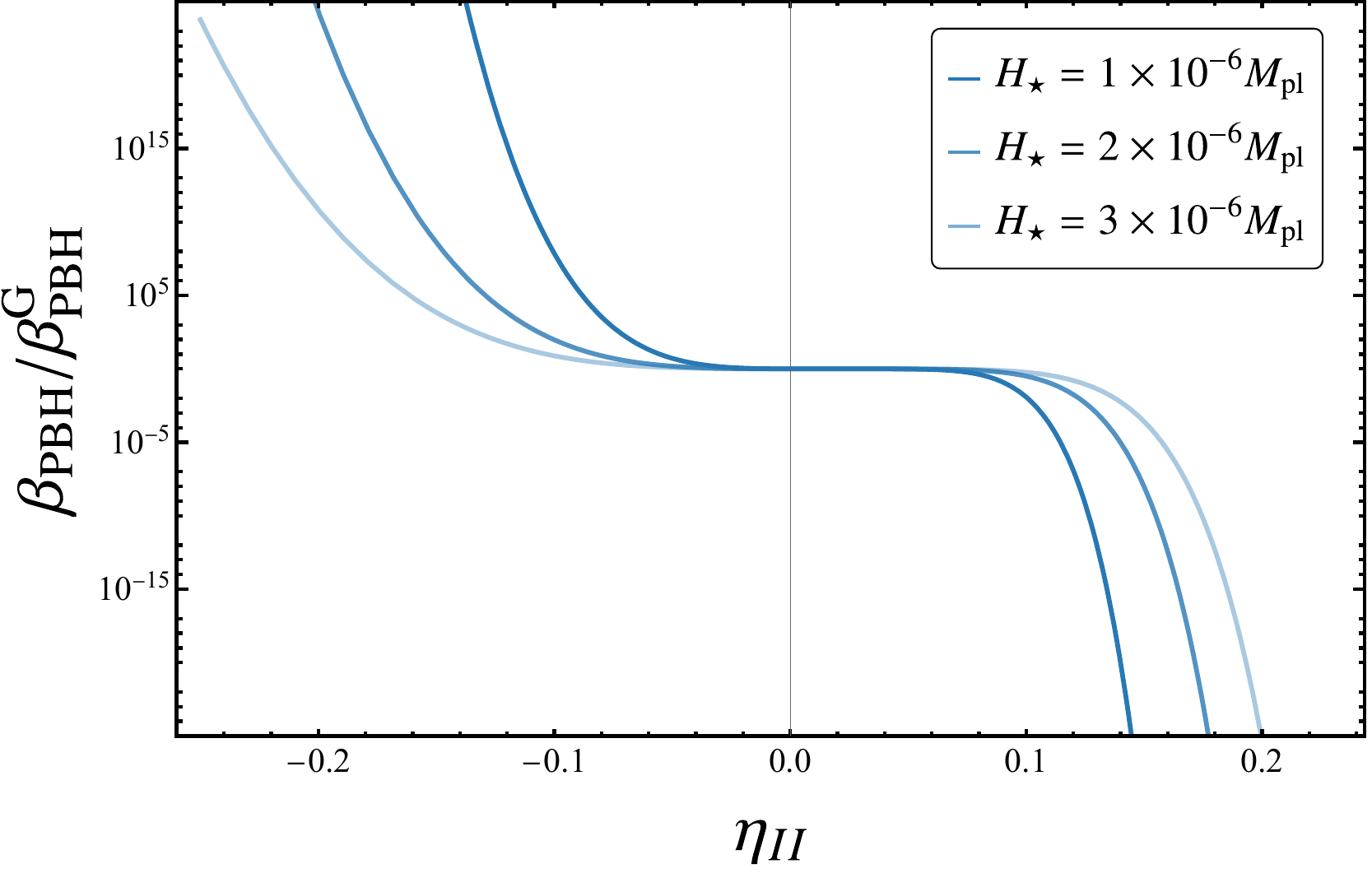}

\caption{\textbf{The ratio $\beta_{\rm PBH}/\beta_{\rm PBH}^{\mathrm{G}}$.}
Top panel: 
For the red dashed (blue solid) lines from dark to light, we take  $\eta_{{II}}=-(+)0.5,-(+)0.3,-(+)0.1$. As a reference, we plot the black solid line with $\eta_{II}=0$ where the $\beta_{\rm PBH}$ overlaps with $\beta_{\rm PBH}^{\mathrm{G}}$.
Bottom panel:
The $\eta_{II}$-dependence of the PBH abundance. 
}
\label{betaratio}
\end{figure}

Let us denote the abundance of PBH that fully includes non-Gaussianity by $\beta_{\rm PBH}=\beta_{\rm PBH}(\delta\phi_{\mathrm{th}})$, while that ignores non-Gaussianity by $\beta_{\rm PBH}^{\mathrm{G}}=\beta_{\rm PBH}(\delta\phi_{\mathrm{th}}^{\mathrm{G}})$.
For definiteness, we fix the curvature perturbation threshold to unity, $\mathcal{R}_{\rm th}=1$ (which corresponds to $\delta_{\rm th}=4/9=0.444\cdots$). The ratio $ \beta_{\rm PBH}/\beta_{\rm PBH}^{\mathrm{G}}$ as a function of $\delta\phi_{\rm th}$ is shown in the upper panel of Fig.~\ref{betaratio}. 
We can see from the figure that the modification of the abundance caused by the nonlinearity grows while the threshold grows. 
Interestingly, for the same absolute value of $\eta_{II}$, the suppression caused by a positive $\eta_{II}$ is always more significant than the enhancement caused by a heavy tail when $\eta_{II}$ is negative. 
This result can be understood from the right-hand side of the second equality in \eqref{th}, i.e., the increase in $\delta\phi_{\rm th}$ is accumulative for $\eta_{II}>0$ while the decrease in $\delta\phi_{\rm th}$ is not for $\eta_{II}<0$. 
The lower panel of Fig.~\ref{betaratio} shows the $\eta_{II}$-dependence of the ratio $ \beta_{\rm PBH}/\beta_{\rm PBH}^{\mathrm{G}}$. One can also see that the suppression is more significant than the enhancement for the same absolute value of $\eta_{II}$.
\section{Conclusion and Discussion}

In this paper, we thoroughly studied a two-stage inflationary model with piecewise quadratic potential. 
Allowing the discontinuity in both the first and second derivatives of the potential, we explored distinct features in the primordial curvature power spectrum and the non-Gaussianity in the probability distribution function (PDF) of the curvature perturbation. 

We obtained an analytic expression for the curvature perturbation power spectrum for the general initial condition by solving the perturbation equation until the end of inflation.
Then for the modes $k<k_\star$ with the adiabatic vacuum initial condition, we obtained a general expression valid up to next-to-leading order corrections in $k/k_\star$. 
It is known that the amplitude of the curvature perturbation that crosses the horizon before the transition, namely, $k<k_\star$ where $k_\star$ is the comoving wavenumber that crosses the horizon at the transition, can be strongly enhanced on superhorizon scales. From the expression we obtained, we clarified the condition under which the commonly observed $k^4$ growth appears and the fine-tuned condition under which the maximum growth rate of $k^5(\ln k)^2$ is realized~\cite{Carrilho:2019oqg}.

The amplitude of the small scale power spectrum (i.e., at $k>k_\star$) relative to the large scale one is found to be proportional to $(\alpha-1)^{-2}$ where $\alpha$ is a function of the first and second derivatives of the potential at both stages, first introduced in this paper, and $\alpha-1$ is a measure of inflaton's attractor component at the beginning of the second stage. 
If $\alpha=1$, the attractor component at the second stage vanishes, and the enhancement becomes infinite. 
Of course, this divergence is an artifact due to our approximation, which will disappear in realistic situations. 

We also observed that there often appears a dip in the power spectrum, and its position is sensitive to the values of the second derivative of the potential (i.e., the $\eta_V$ parameter) at both stages. 
The dip disappears when $\alpha>1$, which corresponds to the case where the potential at the second stage is convex and the inflaton does not reach the attractor phase when it passes the minimum of the potential.


After clarifying many features of the curvature perturbation power spectrum, we discussed the non-Gaussianity.
Adopting the $\delta  N$ formalism, we obtained the formula for the comoving curvature perturbation $\mathcal{R}$ as a function of the field perturbation $\delta\phi$ in flat slicing. 
The formula was originally derived in \cite{Pi:2022ysn}. Here we performed a more detailed analysis of non-Gaussian properties.
Specifically, we presented approximate analytical expressions for the probability distribution function of $\mathcal{R}$ in three cases: (1) the modes leaving the horizon well before the transition, (2) the modes exiting just before the transition, and (3) the modes exiting after the transition. 
In doing so, we clearly identified for the first time the scale dependence in the shape of the PDF as the motion of the inflaton varies in time from the initial slow-roll stage to the non-slow-roll stage right after transition and settles down to the final slow-roll stage. 
In particular, assuming Gaussian $\delta\phi$, we demonstrated that the modes exiting near the transition produce an asymmetric tail for the PDF of $\mathcal{R}$, with the negative tail determined by $\eta_I$ and the positive tail determined by $\eta_{II}$.

Finally, we discussed the dependence of the primordial black hole (PBH) abundance on the shape of the PDF tails.
In our model, we found that the decisive parameter is the second derivative of the potential at the second stage, $\eta_{II}$.
We explored the $\eta_{II}$-dependence using the Press-Schetcher formalism and found that the PBH formation is strongly suppressed for $\eta_{II}>0$ while enhanced for $\eta_{II}<0$, under the assumption that the RMS variation of the field perturbation $\sigma_{\delta\phi}$ is below the threshold for PBH formation in the Gaussian case.
For the same absolute values of $\eta_{II}$, suppression is found to be more prominent than enhancement.
\acknowledgments
We would like to thank Yi-Fu Cai, Xin-Chen He, Shi Pi, Xiaoding Wang, and Ying-Li Zhang for useful discussions.
This work is supported in part by JSPS KAKENHI Grant Nos. JP20H05853 and JP24K00624, National Key R\&D Program of China (2021YFC2203100), by NSFC (12261131497), by CAS young interdisciplinary innovation team (JCTD-2022-20), by 111 Project (B23042), by Fundamental Research Funds for Central Universities, and by CSC Innovation Talent Funds. Kavli IPMU is supported by World Premier International Research Center Initiative (WPI), MEXT, Japan.
XW and XHM gratefully acknowledge the hospitality and support of the Kavli Institute of Physics and Mathematics of the Universe (Kavli IPMU), the University of Tokyo, during their visit when the work has been done. 
We would also like to acknowledge the valuable input and discussions from the IBS CTPU-CGA, Tokyo Tech, USTC 2024 summer school held at Tateyama, Dynamics of Primordial Black Hole Formation II (DPBHF2) workshop held at Nagoya University, and Special Seminar at the University of Science and Technology of China, which greatly contributed to the completion of this work.

\appendix
\bibliographystyle{JHEP}
\bibliography{biblio.bib}
\end{document}